\documentclass[useAMS,usenatbib]{mn2e}

\title[Evolution of the clustering of LRGs]{The 2dF-SDSS LRG and QSO survey: Evolution of the clustering of Luminous Red Galaxies since {\it z} = 0.6}
\author[Wake  et al.]{ \parbox{\textwidth}{
David A. Wake$^1$, 
Ravi K. Sheth$^2$, 
Robert C. Nichol$^3$, 
Carlton M. Baugh$^1$, 
Joss Bland-Hawthorn$^4$, 
Russell Cannon$^5$, 
Matthew Colless$^5$, 
Warrick J. Couch$^6$,  
Scott M. Croom$^4$,
Roberto De Propris$^{7}$, 
Michael J. Drinkwater$^{8}$, 
Alastair C. Edge$^1$, 
Jon Loveday$^{9}$, 
Tsz Yan Lam$^2$, 
Kevin A. Pimbblet$^8$, 
Isaac G. Roseboom$^{9}$,
Nicholas P. Ross$^{1,10}$, 
Donald P. Schneider$^{10}$, 
Tom Shanks$^1$, 
Robert G. Sharp$^5$.}\\\\
$^{1}$Dept. of Physics, Durham University, South Road, Durham, DH1 3LE, UK\\
$^{2}$Dept. of Physics and Astronomy, University of Pennsylvania, 209 South 33rd Street, Philadelphia, PA 19104, USA\\
$^{3}$Institute of Cosmology and Gravitation, University of Portsmouth, Portsmouth, PO1 2EG, UK\\
$^{4}$School of Physics, University of Sydney, NSW 2006, Australia \\
$^5$Anglo--Australian Observatory, PO Box 296, NSW 1710, Australia\\ 
$^6$Centre for Astrophysics and Supercomputing, Swinburne University of Technology, Hawthorn, VIC 3122, Australia\\
$^{7}$Cerro Tololo Inter-American Observatory, La Serena, Chile\\
$^{8}$Dept. of Physics, University of Queensland, Brisbane, Queensland, QLD 4072, Australia\\
$^{9}$Astronomy Center, University of Sussex, Falmer, Brighton, BN1 9QH, UK\\
$^{10}$Dept. of Astronomy and Astrophysics, The Pennsylvania State University, 525 Davey Laboratory, University Park, PA 16802, USA\\}
\begin{document}

\date{}

\pagerange{\pageref{firstpage}--\pageref{lastpage}} \pubyear{}

\maketitle

\label{firstpage}

\begin{abstract}
We present an analysis of the small--to--intermediate scale clustering of samples of Luminous Red Galaxies (LRGs) from the Sloan Digital Sky Survey (SDSS) and the 2dF-SDSS LRG and QSO (2SLAQ) survey carefully matched to have the same rest-frame colours and luminosity. We study the spatial two--point auto-correlation function in both redshift-space ($\xi(s)$) and real--space ($\xi(r)$) of a combined sample of over 10,000 LRGs, which represent the most massive galaxies in the universe with stellar masses $>10^{11}h^{-1}M_{\odot}$ and space densities $\simeq 10^{-4}h^3$Mpc$^{-3}$. 
We find no significant evolution in the amplitude ($r_0$) of the correlation function with redshift, but do see a slight decrease in the slope ($\gamma$) with increasing redshift over $0.19<z<0.55$ and scales of $0.32<r<32h^{-1}$Mpc. 
We compare our measurements with the predicted evolution of dark matter clustering and use the halo model to interpret our results. We find that our clustering measurements are inconsistent ($>99.9\%$ significance) with a passive model whereby the LRGs do not merge with one another; a model with a merger rate of 7.5$\pm$2.3\% from $z=0.55$ to $z=0.19$ (i.e. an average rate of 2.4\% Gyr$^{-1}$) provides a better fit to our observations.  Our clustering and number density measurements are consistent with the hypothesis that the merged LRGs were originally central galaxies in different haloes which, following the merger of these haloes, merged to create a single Brightest Cluster Galaxy (BCG). 
In addition, we show that the small scale clustering signal
constrains the scatter in halo merger histories.  When combined with
measurements of the luminosity function, our results suggest that this
scatter is sub-Poisson. While this is a generic prediction of hierarchical models, it has not been tested before.  
\end{abstract}


\section{Introduction}

In recent years, the evolution of massive galaxies in the universe has received much attention because of the possible tension between observations of the abundance and clustering of such galaxies, as a function of redshift, and predictions from popular hierarchical models of galaxy evolution. Naively, in a Cold Dark Matter (CDM) dominated universe, one would expect the most massive galaxies to form last through the hierarchical merging of smaller galaxies. This behaviour is illustrated in the recent high resolution simulations of \citet{2006MNRAS.366..499D}, which include the latest semi--analytical formalism and account for feedback from active galactic nuclei (AGNs). In these simulations, the stars in the most massive galaxies are formed at high redshifts, but their stellar mass is only assembled into a single system at relatively late times through ``dry mergers", i.e., major mergers of gas--poor galaxies with little or no associated star--formation. For example, in Figures 4 and 5 of \citet{2006MNRAS.366..499D}, the simulations shows that for low redshift elliptical galaxies, with masses $>10^{11}{\rm M_{\odot}}$, 80\% of their stars are formed at a median redshift of $z\simeq2.5$, but 80\% of the stellar mass is only put in place by $z\simeq0.3$. Likewise, the simulations show that galaxies with masses $>10^{11}{\rm M_{\odot}}$ have multiple large progenitors and cannot be formed through a single major merger of two large galaxies. 

These recent AGN--feedback models of galaxy evolution \citep[see also][]{2006MNRAS.365...11C,2006MNRAS.370..645B,2006ApJS..163....1H} solve the apparent inconsistency of the old ages of stars in massive galaxies (both in and outside galaxy clusters) and the late assembly of such galaxies in a $\Lambda$--dominated CDM universe. However, they appear to be in conflict with recent observations of the luminosity function and clustering of massive ellipticals as a function of redshift. For example, \citet{2006MNRAS.372..537W} showed that the lack of evolution of the luminosity function of Luminous Red Galaxies \citep[LRGs; as defined in][]{2001AJ....122.2267E,2006MNRAS.372..425C} put an upper limit on the amount of allowed evolution in these massive galaxies, i.e., at least half of the LRGs at low redshift ($z\sim0.2$) must already have been well assembled (with more than half their stellar mass in place) by $z\sim 0.6$. This is in excellent agreement with other luminosity function studies.  For example, \citet{2007ApJ...654..858B} used data from the NOAO Deep Wide-Field Survey (NDWFS) and the Spitzer IRAC Shallow Survey to show that ``$\simeq80\%$ of the stellar mass contained within today's $4L^*$ red galaxies was already in place at $z=0.7$". These observational constraints are barely consistent with the semi-analytical CDM simulations discussed above. 

The clustering of massive ellipticals provides an additional test of the models. \citet{2006ApJ...644...54M} argue that the small--scale clustering of LRGs from the Sloan Digital Sky Survey \citep[SDSS;][]{2000AJ....120.1579Y} suggest that LRG--LRG mergers (i.e. a major merger of two equally massive systems) were not important for the mass growth of LRGs below $z=0.36$.  More recently, \citet{2007arXiv0708.3240M} used the LRG--galaxy cross--correlation function to study the small--scale clustering of LRGs at $z\sim0.25$ and concluded that LRGs grow in stellar mass at most by $\simeq10\%$ between $0.1<z<1$ (or approximately half the age of the Universe). \citet{2007ApJ...655L..69W} interpretted the evolution in the clustering of luminous red galaxies in the NDWFS using the halo model - they argue that a third of all satellite galaxies (in a halo) disappear over the redshift range $0.5<z<0.9$.  Since the satellite fraction in their models is of order 20\%, only about 7\% of the galaxies have merged.  However, if these mergers increase the stellar mass of the central object, then this increase can be 25\% or even larger.  \citet{2006ApJ...652..270B} report rapid evolution in the stellar mass of red galaxies since $z\simeq1$. This apparent discrepancy is probably due to the differences in the luminosity distributions of the samples, as it is known in clusters that most of the evolution on the so--called ``red sequence" is at magnitudes fainter than $L^*$ \citep[see][and references therein]{2006MNRAS.366..499D,2007ApJ...661...95S}.

In this paper, we expand our earlier study of the evolution of the LRG luminosity function \citep[][Paper I]{2006MNRAS.372..537W} to include an investigation of the two--point auto-correlation function of these galaxies. The key difference of this work to that in the literature is the combination of two large samples of LRGs from the SDSS and the 2dF--SDSS LRG and QSO (2SLAQ) survey \citep{2006MNRAS.372..425C}. As in Paper I, we are careful to ensure the colour selection of LRGs is consistent between these two surveys, thus allowing a study of this unique population of massive ellipticals across the redshift range of $0.15<z<0.6$. In addition, this paper uses the halo model to understand the evolution of the clustering of galaxies and constrain the merger rates of LRGs. Although the logic is similar to the \citet{2007ApJ...655L..69W} analysis of NDWFS, our halo model is entirely analytic, rather than entirely simulation-based. Our analysis is complementary to that of \citet{2007MNRAS.381..573R} who study the redshift space correlation function  of the 2SLAQ sample, binned in pair
separation parallel and perpendicular to the line of sight, and fit both biasing and cosmological parameters to this data. \citet{2007MNRAS.381..573R} conclude that ``LRGs have a constant space density and their clustering evolves purely under gravity", which is consistent with the results of Paper I. Here we wish to test if this conclusion remains true under a more precise comparison of the evolution of the correlation function of LRGs where we accurately account for the changing definition of an LRG with redshift.. 

In Section \ref{sec:data}, we describe the SDSS and 2SLAQ data used in this paper, while in Section \ref{sec:samp} we provide details of the sample selection used to ensure a consistent definition of an LRG across the two samples.  In Section \ref{sec:2ptcalc}, we present our measurements of the two--point correlation function in both real and redshift--space. Section \ref{sec:halo} presents a halo model analysis of our measurements and discuss constraining the merger rate in the halo model framework in Section \ref{sec:HODmerge}. We discuss our findings in the context of recent work in Section \ref{sec:comp} and conclude in Section \ref{sec:discuss}. Throughout this paper, we assume a flat $\Lambda$--dominated cosmology with $\Omega_m=0.27$, $H_0=70 $km s$^{-1} $Mpc$^{-1}$, and $\sigma_8=0.8$ unless otherwise stated. 

\section{Data }
\label{sec:data}
We present in this paper an analysis of galaxies taken from both the
Sloan Digital Sky Survey (SDSS) and the 2dF--SDSS LRG and QSO (2SLAQ) survey. 
The SDSS survey contains 2 main spectroscopic galaxy data sets: the MAIN sample and the LRG sample. The MAIN sample consists of all galaxies with a Galactic extinction corrected petrosian {\it r} magnitude $r_{pet} <$ 17.77; this results in a median redshift of $\sim$ 0.1 \citep{2002AJ....124.1810S}. The LRG sample uses a series of colour and magnitude cuts with the aim of selecting LRGs out to {\it z} $\sim$ 0.5 (see Eisenstein et al. 2001 for details of this sample). Here we only consider the Cut I LRG sample, which has a magnitude limit $r_{pet} <$ 19.2 and is designed to select a pseudo volume--limited sample
of LRGs, with $M_r \le -21.8$ and $0.15 < z < 0.35$. At low redshift there is considerable overlap between the MAIN and LRG samples. We select these two samples of galaxies from the SDSS Data Release 5 \citep{2007ApJS..172..634A}.

The 2SLAQ LRG survey was designed to extend the SDSS LRG sample to $z \sim 0.7$. The LRGs were again selected with colour and magnitude cuts using the SDSS imaging. Spectra were obtained with the 2dF spectrograph on the Anglo-Australian Telescope. Full details of the selection and observations are given in \citet{2006MNRAS.372..425C}. The final LRG sample contains over 11000 LRG
redshifts, covering 180 ${\rm deg^2}$ of SDSS imaging data with over 90\% of these galaxies within
the redshift range $0.45<z<0.7$. The targeted LRGs were split into three
subsamples as detailed in \citet{2006MNRAS.372..425C}, with the primary sample
(Sample 8) accounting for two thirds of these. We only focus on Sample
8 in this paper due to its high completeness and uniform selection.
The overall success rate of obtaining redshifts from the 2dF spectra
for Sample 8 LRGs is 95\%, while the centers of the 2dF fields were
spaced by 1.2$^{\circ}$, resulting in an overall redshift completeness
of sample 8 LRG targets of $\sim$75\% across the whole survey area
(Paper I).

Although the SDSS magnitude system \citep{1996AJ....111.1748F} was designed to be on the AB scale
\citep{1983ApJ...266..713O}, the final calibration has differences
from the proposed values by a few percent. We have applied the
corrections m$_{AB}$ = m$_{SDSS} + [-0.036, 0.012, 0.010, 0.028,
0.040]$ for $u$, $g$, $r$, $i$, $z$ respectively (Eisenstein, priv.
comm.).  All magnitudes and colours presented throughout this paper
are corrected for Galactic extinction \citep{1998ApJ...500..525S}.

\section{Matching Samples}
\label{sec:samp}

\begin{figure}
\vspace{9.0cm}
\includegraphics{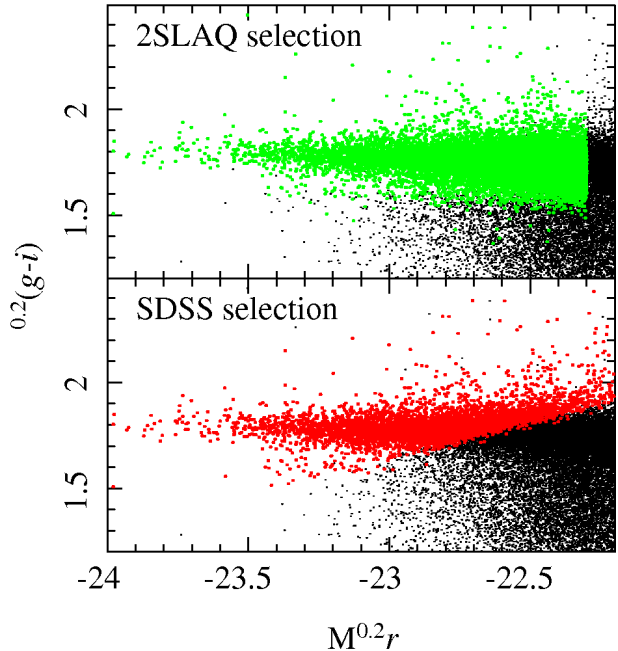}

\caption{\label{fig:sel} The $^{0.2}(g-i)$ versus M$_{^{0.2}r}$ colour magnitude relation for SDSS main galaxies with $0.15 < z < 0.21$, all K+e corrected to $z$ = 0.2. The black points in each panel show the whole sample. The top panel shows those galaxies that are selected to be in the 2SLAQ selection matched sample (see text) when K+e corrected to $z$ = 0.55 (green points). The bottom panel shows those galaxies that would be selected in the the SDSS selection matched sample (see text) when K+e corrected to $z$ = 0.2 (red points).}
\end{figure}

Different techniques were employed to select LRGs in the SDSS and the 2SLAQ survey, resulting in intrinsic differences between the properties of the LRGs in each sample (Figure \ref{fig:sel}). In particular, the magnitude dependent colour cut used in the SDSS selection results in only the very reddest galaxies being included in the SDSS LRG sample at fainter magnitudes. Therefore, if we wish to make a meaningful comparison of the evolution of LRGs with redshift we must make additional colour and magnitude cuts to ensure that we exactly match the samples from the two surveys. 

Following Paper I, we assume that the evolution of the LRGs stellar populations can be approximated by simple passive ageing. We therefore use the same models as Paper I to generate K+e corrections which are used to correct the observed magnitudes of each sample to a common frame. Paper I demonstrated that these models do not perfectly describe the colour evolution of the LRGs because of inadequacies in the stellar population synthesis models. To minimize the magnitude of these corrections, Paper I restricted their LRG samples to tight redshift ranges at approximately {\it z} = 0.2 and {\it z} = 0.55 where the {\it u,g} and {\it r} filters approximately map onto the {\it g, r} and {\it i} filters respectively. These same redshift cuts are again applied to the samples used herein.

In this paper we take two approaches to matching the selection between these two redshifts. In the first we follow the procedure of Paper I. We take all the SDSS LRGs with 0.17 $< z <$ 0.24 and K+e correct their magnitudes to both z=0.2 and z=0.55. We then apply the SDSS selection criteria using the {\it z} = 0.2 magnitudes and the 2SLAQ selection criteria using the {\it z} = 0.55 magnitudes. We then execute the same procedure on the 2SLAQ LRGs within 0.5 $< z <$ 0.6. We note that since the 2SLAQ selection is significantly bluer in the rest-frame than the SDSS selection; it is the application of SDSS selection cuts that is removing the majority of the LRGs removed from each sample by this procedure. We will therefore describe these samples as the SDSS selection matched samples.

Our second approach makes use of the MAIN galaxy sample from the SDSS rather than just the LRG sample, although there is considerable intersection over the redshift range we are considering here. We limit the MAIN galaxies to 0.15 $< z <$ 0.21 and then apply our K+e corrections to correct to both {\it z} = 0.2 and {\it z} = 0.55. For the galaxies at {\it z} = 0.21 the $r_{pet}$ = 17.77 magnitude limit of the MAIN galaxy sample corresponds to M$_{^{0.2}r}$ = -22.3.  M$_{^{0.2}r}$ is calculated by determining the apparent magnitude the galaxy would have at {\it z} = 0.2 in the SDSS $r$-band filter using our assumed K+e corrections, and is then converted to an absolute magnitude using the distance modulus without the use of any further K or evolutionary corrections.  The M$_{^{0.2}r}$ = -22.3 is only 0.3 magnitudes brighter than the limit of the 2SLAQ sample when K+e corrected to this redshift. Since the MAIN sample contains galaxies of all colours we can generate a sample matching the 2SLAQ selection and we only need limit the 2SLAQ sample by this M$_{^{0.2}r}$ cut. There is, however, an additional complication. As shown in Paper I the errors on the photometry at the faint magnitudes of 2SLAQ result in a large scatter of objects across the colour and magnitude selection boundaries. To mimic this effect we measure the magnitude error distributions of the 2SLAQ galaxies as a function of magnitude and modify the magnitudes of the SDSS galaxies randomly following this error distribution. We then apply the 2SLAQ selection criteria to both samples K+e corrected to {\it z} = 0.55 along with a cut at M$_{^{0.2}r}$ = -22.4. This slightly brighter cut than the M$_{^{0.2}r}$ = -22.3 limit allows the inclusion fainter galaxies which are being scattered into the selection region by the application of the 2SLAQ photometric errors mimicking the effect present in the 2SLAQ data. We will refer to these samples as the 2SLAQ selection matched samples.

We are unable to account for the effect of the photometric errors on the selection in the SDSS selected LRG sample as we don't have galaxies in that sample which are fainter or bluer than the LRGs. In Paper I we corrected the LF at {\it z} = 0.55 for this sample using a sub-region that had deeper photometry. We are unable to apply such a correction in this work since the significantly smaller area ($\sim$1/3 of the total) of this sub-region would result in a very poor measurement of the clustering and render any correction highly unreliable. The smaller area was not a problem for the LF measurement since the region of the LF most affected was the faint end were the galaxies were most numerous. The correction was also only required for a subsection of the LF, which one could always choose to disregard, whereas it would affect the entire correlation function.
For this reason, when making direct evolutionary comparisons between redshifts, we will focus on the 2SLAQ selection matched samples.
Table \ref{tab:samp} gives the number of galaxies in each sample and Figure \ref{fig:sel} illustrates the difference between the two selection criteria.

\begin{table}
  \begin{center}
    \caption{\label{tab:samp} The redshift range, selection and number of galaxies in each sample defined in the text.}
    \begin{tabular}{c  c  c  c} 
      \multicolumn{1}{c}{Sample} &
      \multicolumn{1}{c}{Redshift} &
      \multicolumn{1}{c}{Selection} &
      \multicolumn{1}{c}{Number} \\
      \hline \hline 
       1  & 0.17 $< z <$ 0.24  & SDSS  &  9,912\\
       2  & 0.5 $< z <$ 0.6  & SDSS  &  1,239\\
       3  & 0.15 $< z <$ 0.21  & 2SLAQ  & 11,350\\
       4  & 0.5 $< z <$ 0.6  & 2SLAQ  & 2,814\\
      \hline
    \end{tabular}
  \end{center}
\end{table}

\begin{figure*}
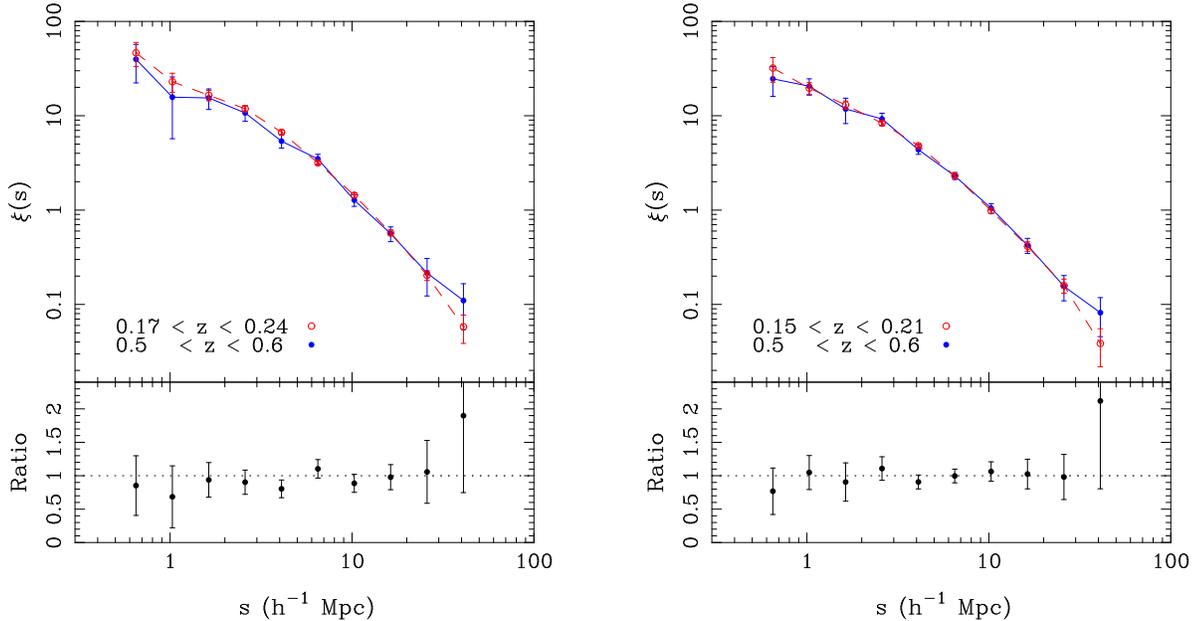

\vspace{8.5cm}
\includegraphics{2ptLRG_psis_rat_paper.ps}
\includegraphics{2ptLRG_main_psis_rat_paper.ps}

\caption{\label{fig:zspace} The redshift space 2pt-correlation functions at $z\sim0.2$ (red open circles) and $z\sim0.55$ (blue filled circles) and their ratio ($z \sim$ 0.55/$ z \sim$ 0.2) for the SDSS selection matched (left) and for the 2SLAQ selection matched (right) samples.}
\end{figure*}

\section{The measured 2pt-correlation function}
\label{sec:2ptcalc}

The 2pt-correlation function, $\xi(r)$, is defined as the excess probability above Poisson of finding an object at a separation $r$ from another object. This is calculated by comparing the number of pairs as a function of scale in our galaxy catalogues, with the number in a random catalogue, which covers the same volume as our data. We make this measurement using the \citet{1993ApJ...412...64L} estimator,
\begin{equation}
	\xi = \frac{1}{RR}\left[DD\left(\frac{n_R}{n_D}\right)^2 - 2DR\left(\frac{n_R}{n_D}\right) + RR\right],
\end{equation}
where DD, DR and RR are data-data, data-random and random-random pair counts respectively, and $n_D$ and $n_R$ are number of galaxies in the data and random catalogues.  

\begin{figure*}
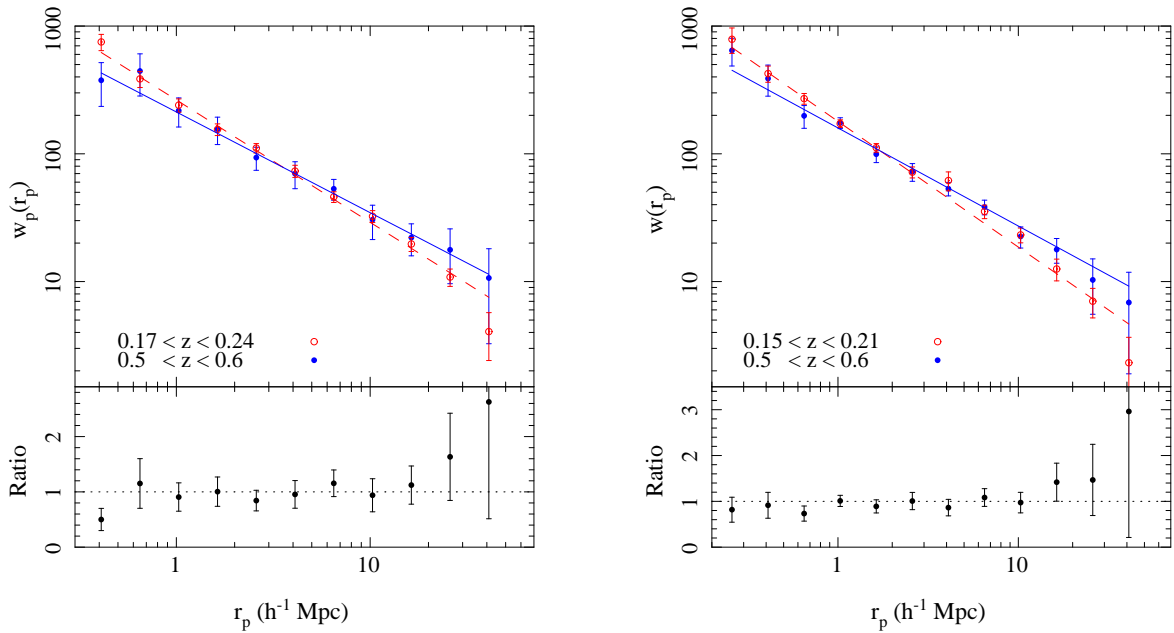

\vspace{8.5cm}
\includegraphics{2ptLRG_wrp_rat_paper.ps}
\includegraphics{2ptLRG_main_wrp_rat_paper.ps}

\caption{\label{fig:wrp} The projected 2pt-correlation functions at $z\sim0.2$ (red open circles) and $z\sim0.55$ (blue filled circles) and their ratio ($z \sim$ 0.55/$ z \sim$ 0.2) for the SDSS selection matched (left) and for the 2SLAQ selection matched (right) samples. The lines show power law fits on scales $0.32 < r_p < 32 h^{-1}$ Mpc.}
\end{figure*}

\subsection{Incompleteness corrections and error estimates}
When making the 2pt-correlation function  measurement in our samples we must account for the varying completeness across our surveys. For both the SDSS and 2SLAQ we separate the galaxies into unique regions based on the positions of the overlapping spectroscopic plates. For the SDSS survey we use the regions defined in the SDSS Catalogue Archive Server \citep[see][for details.]{2007ApJS..172..634A}. For 2SLAQ we use regions defined using the angular mask constructed using repeated runs of the 2dF-configure software (see Paper I for details).  Within each region we determine the number of targets with reliable redshifts ($N_R$) and the number of targets that could have been observed ($N_T$). The completeness in each region is then defined as the ratio of these ($N_R/N_T$). Regions with completeness below 65\% are removed. 

To correct for the remaining incompleteness we wish to assign a weight $\ge$ 1 to all the galaxies that have a reliable redshift. We begin by assigning each target galaxy a weight equal to the inverse of the completeness of the region in which it lies. For those that do not have a redshift the weight is redistributed to its three nearest neighbours. This maintains some of the spatial information, although on the smallest scales where fibre collisions become important the clustering signal is likely to be underestimated. The weights of the galaxies with redshifts in a given region are then renormalised so that the mean weight in that region is as it was before the redistribution, $i.e.$ the inverse of the completeness. 

An alternative approach, often used to correct for incompleteness, is to reduce the number of random points in regions with low completeness. We do not do this for two reasons. Firstly, by having regions with lower numbers of random points we will be unnecessarily increasing the noise in these regions. Secondly, and more importantly, unlike in the SDSS, the spectroscopic plates in 2SLAQ were evenly spaced with no allowance made for the variation of the target space density. This means that regions with a high target density (i.e. highly clustered regions) will be more likely to have a lower completeness. We calculate the completeness in regions defined by the overlapping plates and so by simply reducing the number of random points based on this completeness we would be likely to systematically underestimate the clustering on scales smaller than the given region. We would be preferentially removing the most clustered galaxies and then renormalising the clustering calculated from the remaining {\it less} clustered galaxies by the ratio of the number removed (i.e. the completeness). Since we instead redistribute the weight of the galaxies without redshifts to their nearest neighbours, we are likely to be up weighting other galaxies in the most clustered regions and will therefore be making a better estimate of the true clustering amplitude. 

Nearly all of the completeness regions have annular scales up to 2 degrees which corresponds to 32.6 $h^{-1}$ Mpc at $z = 0.55$ and so this effect is likely to be important over nearly all the scales we consider in this paper. In fact the clustering is $\simeq$ 5\% lower for the 2SLAQ samples when calculated by just reducing the number of randoms. 

We generate random catalogues for each galaxy sample following the angular masks of the surveys with constant space density and 20 times the number of random points as data. The regions around bright stars are removed from both data and random catalogues, as galaxies in these regions are known to have systematically incorrect magnitudes due to poor sky subtraction in SDSS photometric pipeline \citep{2005MNRAS.361.1287M,2006ApJS..162...38A}. Redshifts are assigned to the random catalogues by randomly sampling a polynomial fit to the redshift distribution of each galaxy sample. We note that within the tight redshift ranges of the samples considered here all the samples are approximately volume limited. 

We estimate the errors on our 2pt-correlation function measurements using jackknife re-sampling \citep{2002ApJ...579...48S,2005ApJ...621...22Z}. We split the SDSS area into 40 equal area regions and the 2SLAQ area into 32 equal area regions. We then calculate each 2pt function removing one area at a time to generate a full covariance matrix. Throughout this analysis we measure the pair counts using the KD-tree code in the NTROPY software package \citep{2007arXiv0709.1967G}. 

\subsection{Various clustering estimators}
The peculiar velocities of galaxies generate errors in the distance measurements along the line of sight.  This means that our basic measurement of $\xi$, which is based on redshift distances, is affected by these redshift space distortions.  By separating the clustering signal into contributions perpendicular ($r_p$) and parallel ($\pi$) to the line-of-sight ($\xi(r_p,\pi)$) and then integrating over the $\pi$ direction, one obtains the projected correlation function 
\begin{equation}
\label{eq:wprp}
	w_p(r_p) = 2\int^{\infty}_0 d\pi\,\xi(r_p,\pi)
	        = 2\int^{\infty}_{r_p} \frac{r\,dr\,\xi(r)}{(r^2-r_p^2)^{1/2}}.
\end{equation}
The final expression only involves the real space correlation function $\xi(r)$ showing that $w_p(r_p)$ is not compromised by redshift space distortions \citep{1983ApJ...267..465D}.  One can invert equation~(\ref{eq:wprp}) by interpolating between the binned $w(r_p)$ to yield an estimate of $\xi(r)$ which is free of redshift space distortions \citep{1992MNRAS.258..134S}.  If $\xi(r) = (r/r_0)^{-\gamma}$, then equation~(\ref{eq:wprp}) can be solved analytically \citep{1983ApJ...267..465D}.

\begin{figure*}
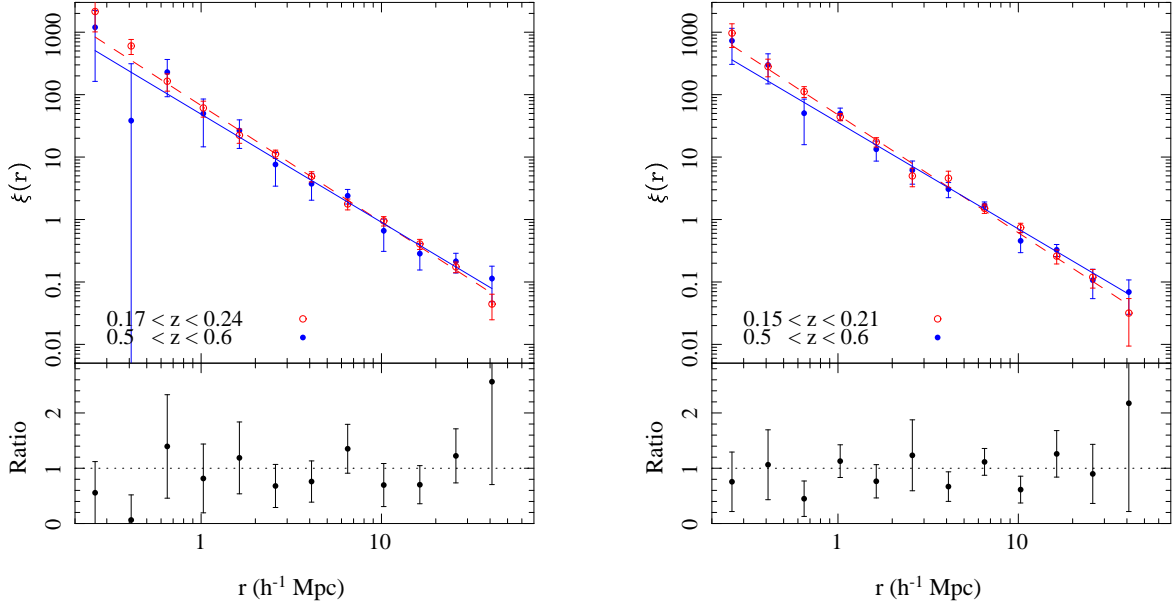

\vspace{8.5cm}
\includegraphics{2ptLRG_psir_rat_paper.ps}
\includegraphics{2ptLRG_main_psir_rat_paper.ps}

\caption{\label{fig:xir} The real space 2pt-correlation functions at $z\sim0.2$ (red open circles) and $z\sim0.55$ (blue filled circles) and their ratio ($z \sim$ 0.55/$ z \sim$ 0.2) for the SDSS selection matched (left) and for the 2SLAQ selection matched (right) samples. The lines show power law fits on scales $0.32 < r < 32 h^{-1}$ Mpc.}
\end{figure*}

\begin{figure*}
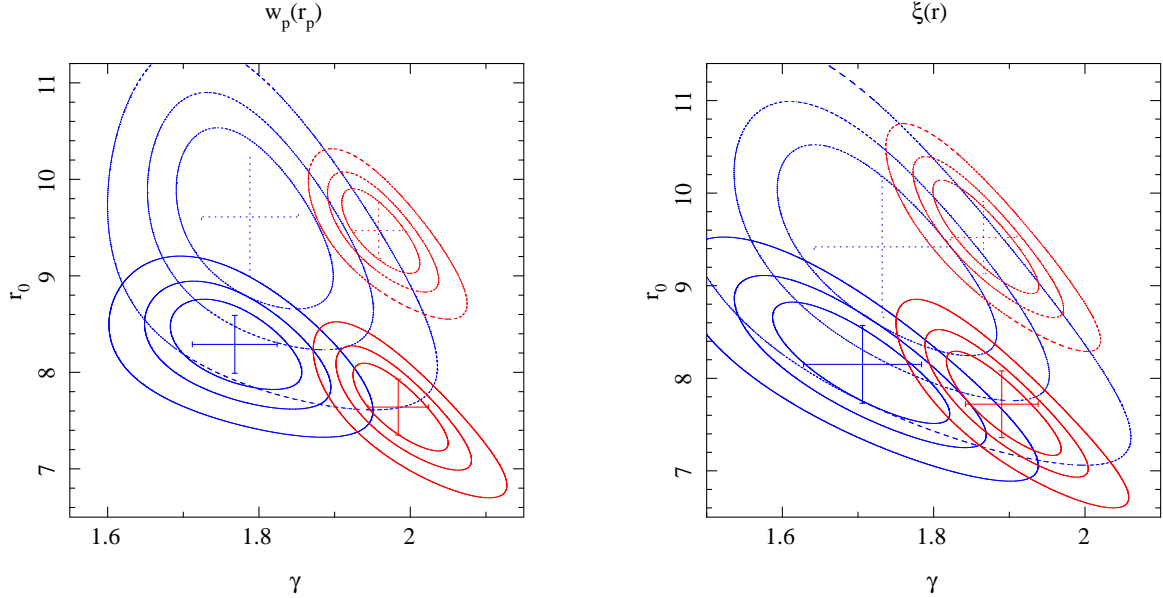

\vspace{8.5cm}
\includegraphics{2ptfit_wrp_cont_sdss_2slaq_paper.ps}
\includegraphics{2ptfit_psir_cont_sdss_2slaq_paper.ps}

\caption{\label{fig:xircont} 68\%, 90\% and 99\% confidence intervals for power law fits on scales $0.32 < r_p < 32 h^{-1}$ Mpc to the projected 2pt-correlation functions (left) and the real space 2pt-correlation functions (right) at $z\sim0.2$ (red) and $z\sim0.55$ (blue) for the SDSS selection matched (dashed lines) and for the 2SLAQ selection matched (solid lines) samples. The error bars show the 1 $\sigma$ errors on the individual parameters.}
\end{figure*}

In practise, one models $w_p$ with the second of the equalities above, but 
measures it using the first.  However, when making the measurement, it is only sensible to integrate out to some maximum $\pi$ because $\xi(r_p,\pi)$ is poorly known on very large scales.  We integrate to 80$h^{-1}$Mpc which appears to give stable results.

\subsection{Observed evolution of clustering}
\label{subsec:rev}

Figures \ref{fig:zspace}--\ref{fig:xir} show $\xi(s)$, $w(r_p)$, and $\xi(r)$ for the four samples described in Section \ref{sec:samp}, along with the ratio of the functions between the two redshifts. Figures \ref{fig:wrp}--\ref{fig:xir} also show the result of fitting power laws over the scales $0.32 < r_p < 32 h^{-1}$ Mpc using the full covariance matrices derived from the jackknife re-sampling technique. We limit the fits to scales greater than $0.32h^{-1}$ Mpc since our weighting scheme does not fully correct for the effect of fibre collisions on smaller scales. Table~\ref{tab:fitpower} provides the best fit values of $r_0$, $\gamma$ and the associated reduced $\chi^2$, with error contours shown in figure \ref{fig:xircont}.
These measurements show that there is very little evolution in the clustering amplitude of LRGs between $z\sim0.55$ and $z\sim0.2$, but there is a marginally significant increase in the slope.

\begin{table*}
  \begin{center}
    \caption{\label{tab:fitpower} Values of the power-law fits and the reduced $\chi^2$ to w$(r_p)$, and $\xi(r)$ in the range $0.32 < r < 32 h^{-1}$ Mpc.}
    \begin{tabular}{c  c  c  c  c  c  c  c} 
      \multicolumn{1}{c}{Selection} &
      \multicolumn{1}{c}{Redshift} &
      \multicolumn{2}{c}{$r_0$ ($h^{-1}$ Mpc)} &
      \multicolumn{2}{c}{$\gamma$} &
      \multicolumn{2}{c}{$\chi^2_{min}$} \\
       \multicolumn{1}{c}{} &
       \multicolumn{1}{c}{} &
      \multicolumn{1}{c}{$w(r_p)$} &
      \multicolumn{1}{c}{$\xi(r)$} &
      \multicolumn{1}{c}{$w(r_p)$} &
      \multicolumn{1}{c}{$\xi(r)$} &
      \multicolumn{1}{c}{$w(r_p)$} &
      \multicolumn{1}{c}{$\xi(r)$} \\
      \hline \hline 
	SDSS & 0.21 & 9.47 $\pm$ 0.29 & 9.52 $\pm$ 0.39 & 1.96 $\pm$ 0.03 & 1.87 $\pm$ 0.04 & 0.76 & 0.54 \\
	SDSS & 0.55 & 9.61 $\pm$ 0.62 & 9.42 $\pm$ 0.76 & 1.79 $\pm$ 0.06 & 1.73 $\pm$ 0.09 & 0.53 & 0.71 \\
	2SLAQ & 0.19 & 7.64 $\pm$ 0.29 & 7.72 $\pm$ 0.36 & 1.98 $\pm$ 0.04 & 1.89 $\pm$ 0.05 & 1.58 & 1.10 \\
	2SLAQ & 0.55 & 8.29 $\pm$ 0.30 & 8.15 $\pm$ 0.42 & 1.77 $\pm$ 0.06 & 1.71 $\pm$ 0.08 &  1.02 & 0.80 \\
      \hline
    \end{tabular}
  \end{center}
\end{table*}

\subsection{Comparison with previous work}
Several previous studies have performed similar analyses to those we present
here; it is important to make a comparison of the results before further investigating the meaning of these measurements. \citet{2005ApJ...621...22Z} present the 2pt-correlation function for three slightly different samples of SDSS LRGs. One of these samples, with $-23.2 < M_g < -21.2$, has an almost identical space density to the $z\sim0.2$ SDSS selection matched sample, although at a higher redshift ($\overline{z} = 0.28$).  The 2pt-correlation functions in redshift, projected and real space for this sample are almost indistinguishable, within the errors, to those presented here. 

\citet{2007MNRAS.381..573R} present measurements of the 2pt-correlation function for the Sample 8 2SLAQ LRGs. This sample is similar to the 0.5 $< z <$ 0.6 2SLAQ selection matched sample although with a larger redshift range and slightly fainter absolute magnitude cut. The power-law fit to $w(r_p)$ in \citet{2007MNRAS.381..573R} has a very similar slope ($\gamma = 1.83\pm0.05$) to that measured here with a lower amplitude ($r_0 = 7.30 \pm 0.34 h^{-1}$ Mpc). This lower amplitude is to be expected as \citet{2007MNRAS.381..573R} include intrinsically fainter galaxies in their sample. To make a more direct comparison we recalculated $w(r_p)$ using a selection almost identical to that used by \citet{2007MNRAS.381..573R}. This produces an almost identical slope ($\gamma = 1.81 \pm 0.03$) but a slightly higher amplitude ($r_0 = 7.85 \pm 0.15 h^{-1}$ Mpc) to that found by \citet{2007MNRAS.381..573R}. This is to be expected, as \citet{2007MNRAS.381..573R} simply reduced the number of randoms points as a function of completeness. As discussed in Section \ref{sec:2ptcalc}, the 2SLAQ data are more likely to be incomplete in the densest regions and so by reducing the number density of random points as a function of completeness they will tend to underestimate the clustering on scales smaller than the regions in which the determine the completeness.

Since we produce almost identical measurements to those presented in \citet{2005ApJ...621...22Z} and \citet{2007MNRAS.381..573R}, with a completely independent analysis and different techniques on largely the same data, we can be confident that our measurements are accurate. We now consider what our measurements imply for our LRG samples.  

The slope of the 2pt-function is known to depend on colour/spectral type: bluer galaxies have a shallower slope \citep[e.g.][]{2002ApJ...571..172Z,2002MNRAS.332..827N}. Could it be that there are more blue galaxies in the $z = 0.55$ samples? We have assumed passive evolution when defining the sample selection, so it seems unlikely that this would include more intrinsically bluer/later-type galaxies at high redshift than at low redshift. We could, however, be scattering more blue galaxies across the selection boundaries at high redshift than at low redshift, for instance, if there were more galaxies populating the blue cloud close to the red sequence at high $z$. If this is the case one might expect to see a difference in the slopes between the 2SLAQ selection matched and SDSS selection matched samples, as the SDSS selection only allows the reddest galaxies to be included at the faintest magnitudes where the scattering is most significant. This is not the case, suggesting that despite the fact that we have selected galaxy populations consistent with purely passive evolution, both dynamically and in terms of their stellar populations, we are in fact seeing some additional evolution in the LRG population. 

\subsection{Comparison with a no-merger model}
\label{sec:nomerge}

If, as suggested in Paper I, the LRGs do not merge with one another, 
then the large scale bias is predicted to evolve as
 $b_{lo} = 1+ (b_{hi}-1) (D_{hi}/D_{lo})$ 
where $D$ is the linear growth factor \citep{1996MNRAS.282..347M,1996ApJ...461L..65F}.  In this case, the ratio of the correlation functions should be  
\begin{equation}
  {\xi_{hi}(r)\over\xi_{lo}(r)} = {b_{hi}^2 D_{hi}^2\over b_{lo}^2 D_{lo}^2}
  = \left({b_{lo} - 1 + D_{hi}/D_{lo}\over b_{lo}}\right)^2
 \label{xiev}
\end{equation}
on large scales.  Note that this differs from the growth of the dark matter clustering strength, because of the factor $(b_{hi}/b_{lo})^2$. Since $D_{hi}/D_{lo}\le 1$, the large scale clustering strength should increase at late times.  For $z_{hi}=0.55$ and $z_{lo}=0.2$ in our chosen cosmology, $D_{hi}/D_{lo}=0.84$.  We will argue below that $b_{hi}/b_{lo} = 2.16/1.91 = 1.13$, so the expected ratio of large scale clustering strengths is 0.9.   

A similar argument can be made for the clustering in redshift-space:  on scales where the \citet{1987MNRAS.227....1K} analysis of redshift space distortions applies, the expected ratio of redshift-space clustering amplitudes is
\begin{equation}
  {\xi_{hi}(s)\over\xi_{lo}(s)} =
   {1 + 2\beta_{hi}/3 + \beta_{hi}^2/5\over 1 + 2\beta_{lo}/3 + \beta_{lo}^2/5}\,
   \left({b_{hi}\,D_{hi}\over b_{lo}\,D_{lo}}\right)^2,
\end{equation}
where $\beta_{lo}\approx \Omega_{lo}^{5/9}/b_{lo}$
and $\beta_{hi}\approx\Omega_{hi}^{5/9}/b_{hi}$.
Again, in the no merger model, the low redshift population is expected 
to be more strongly clustered.
For the two LRG samples studied in the main text, the expected ratio 
is $(1.236/1.211)0.9 = 0.92$.  

Figures \ref{fig:zspace}, \ref{fig:wrp} and \ref{fig:xir} show the ratios of $\xi(s)$, $w_p$ and $\xi(r)$ measured at $z\sim$ 0.55 and $z\sim$ 0.2 which appear to be consistent with little or no evolution. The expected ratios calculated above are inconsistent with the data at the 93\% level for $\xi(s)$ and the 80\% level for $w_p$ on large scales ($r > 3 h^{-1}$ Mpc) where these calculations apply.  Thus, the clustering signals suggest that the low redshift LRG populations are not simply passively evolved versions of the high redshift population, although we are not able to conclusively demonstrate this with the large-scale clustering measurements alone.  In the following sections we model both the evolution of the clustering on all scales and the number density to further constrain the evolution of LRGs.

\section{Halo Model Analysis}\label{sec:halo}
The halo model (see Cooray \& Sheth 2002 for a review) assumes that the galaxy clustering signal encodes information about the Halo Occupation Distribution (HOD) - how the galaxies populate Dark Matter haloes - in particular, how the HOD depends on halo mass. This approach has recently been used to constrain the HODs of galaxies in a number of large data sets.  We apply such a model here to try to gain insight into our LRG populations, how they have evolved, and how well or otherwise this evolution can be described by the passive no-merger model.  Our analysis of the no-merger model has strong similarities to that recently performed by \citet{2007ApJ...655L..69W} and \citet{2007arXiv0712.1643S}.  However, whereas their work was primarily numerical, our analysis shows that the entire discussion can be analytic.  

\subsection{The centre-satellite HOD}
In the halo model, every galaxy is associated with a halo; all haloes are 200 times the background density whatever the mass $M$ of the halo.  Sufficiently massive haloes typically host more than one galaxy.  The halo model we use distinguishes between the central galaxy in a halo, and the others, which are usually called satellites.  This is motivated by simulations \citep[e.g.][]{2004ApJ...609...35K}, and has been a standard assumption of semi-analytic galaxy formation models for many years \citep[e.g.][]{2006RPPh...69.3101B}.  There is now strong observational evidence that the two types of galaxies are indeed rather different, and that the halo model parametrisation of this difference is rather accurate \citep{2007MNRAS.382.1940S}.  

The fraction of haloes of mass $M$ which host centrals is modelled as 
\begin{equation}
\label{eq:Ncen}
	\langle N_c|M\rangle = \exp(-M_{min}/M). 
\end{equation}
Only haloes which host a central may host satellites.  In such haloes, the number of satellites is drawn from a Poisson distribution with mean 
\begin{equation}
 \label{eq:Nsat}
	\langle N_s|M \rangle = (M/M_1)^{\alpha}.
\end{equation}
Thus, the mean number of galaxies in haloes of mass $M$ is 
\begin{equation}
 \label{eq:Ntot}
 \langle N|M\rangle = \langle N_c|M\rangle[1 + \langle N_s|M \rangle], 
\end{equation} 
and the predicted number density of galaxies is 
\begin{equation}
\label{eq:den}
	n_g =  \int dM\, n(M)\, \langle N|M\rangle,
\end{equation}
where $n(M)$ is the halo mass function, for which we use the parametrisation given by Sheth \& Tormen (1999).

We further assume that the satellite galaxies in a halo trace an NFW profile \citep{1996ApJ...462..563N} around the halo centre, and that the haloes are biased tracers of the dark matter distribution.  The halo bias depends on halo mass in a way that can be estimated directly from the halo mass function (Sheth \& Tormen 1999).  With these assumptions the halo model for $\xi(r)$ is completely specified (e.g. Cooray \& Sheth 2002).  We then calculate $w(r_p)$ from $\xi$ using the second of equations~(\ref{eq:wprp}).

\begin{figure*}
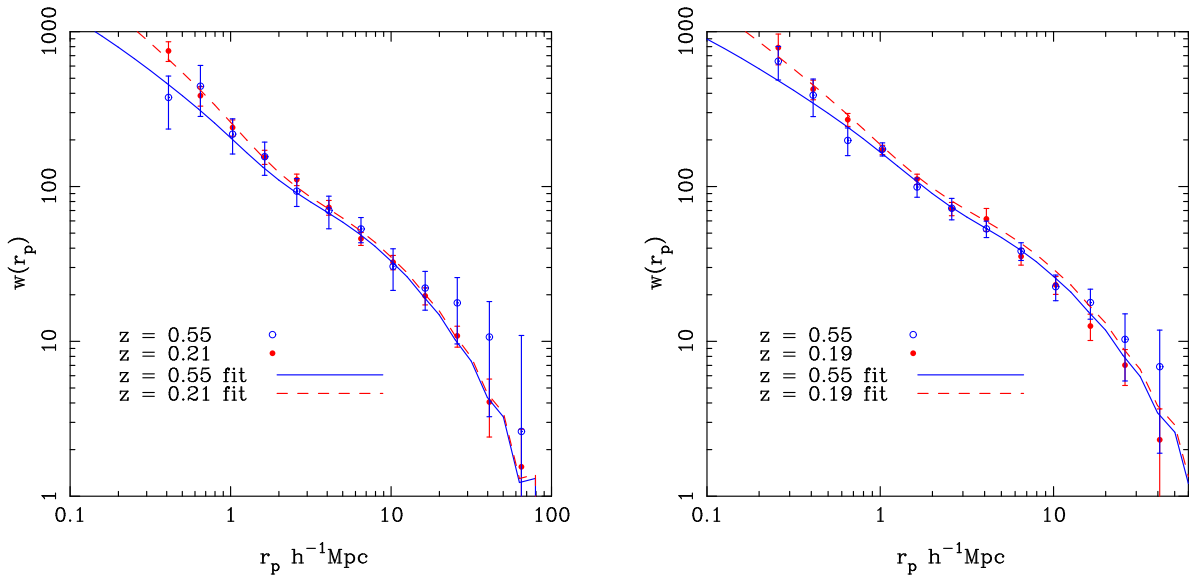

\vspace{8.5cm}
\includegraphics{hodplot_sdss_paper.ps}
\includegraphics{hodplot_paper.ps}

\caption{\label{fig:wphod}  HOD fits on scales $0.32 < r_p < 50 h^{-1}$ Mpc to the projected 2pt-correlation functions at $z\sim0.2$ (red) and $z\sim0.55$ (blue) for the SDSS selection matched (left) and for the 2SLAQ selection matched (right) samples. }
\end{figure*}

In addition to $\xi$, we are interested in the satellite fraction, 
\begin{equation}
 F_{sat} = \int dM\, n(M)\,\langle N_c|M\rangle\,\langle N_s|M\rangle/n_g.  
\end{equation} 
and two measures of the typical masses of LRG host haloes: 
an effective halo mass 
\begin{equation}
\label{eq:Meff}
	M_{eff} =  \int dM\, M\, n(M)\, \langle N|M\rangle/n_g,
\end{equation}
and the average linear bias factor 
\begin{equation}
\label{eq:blin}
	b_{g} =  \int dM\, n(M)\, b(M)\,\langle N|M\rangle/n_g,
\end{equation}
where $b(M)$ is the halo bias.

Our notation is intended to make explicit the fact that the mean number density of central-satellite pairs from such haloes is $n(M)\,\langle N_c|M\rangle\,\langle N_s|M \rangle$, and the mean number density of distinct satellite-satellite pairs is $n(M)\,\langle N_c|M\rangle\,\langle N_s|M \rangle^2/2$ (because we are assuming the satellite counts are Poisson).  

For completeness, our model for the real-space 2-point function is 
\begin{equation}
 \xi(r) = 1+\xi_{cs}(r) + 1+\xi_{ss}(r) + \xi_{2h}(r)
\end{equation}
where 
\begin{eqnarray}
1+\xi_{cs}(r) &=& \int dM\, {n(M)\langle N_c|M\rangle\over n_g}\, \langle N_s|M\rangle\, {\rho(r|M)\over n_gM} \\
1+\xi_{ss}(r) &=& \int dM\, {n(M)\langle N_c|M\rangle \over n_g}\, \frac{\langle N_s|M\rangle^2}{2}\, {\lambda(r|M)\over n_gM^2} 
\end{eqnarray}
and 
\begin{equation}
\xi_{2h}(r) = \int \frac{dk}{k}\,\frac{k^3P_{2h}(k)}{2\pi^2}
\end{equation}
with 
\begin{eqnarray}
 P_{2h}(k) &=& b_g(k)^2\,P_{\rm Lin}(k), \qquad {\rm where}\\
  b_g(k) &=& \int dM {n(M)\over n_g}\, b(M)\,
 \langle N_c|M\rangle\Bigl[1 + \langle N_s|M\rangle u(k|M)\Bigr].\nonumber
\end{eqnarray}
In the expressions above,
 $\rho(r|M)$ is the density profile of haloes of mass $M$,
 $\lambda(r|M)$ denotes the convolution of two such profiles, 
 $u(k|M)$ is the Fourier transform of $\rho(r|M)/M$, 
 and $P_{\rm Lin}(k)$ denotes the linear theory power spectrum.  
In practise, we usually approximate $b_g(k)$ by its value $b_{g}$ 
at $k=0$ (equation~\ref{eq:blin}).  
All these quantities, along with the mass function $n(M)$ and bias 
factor $b(M)$, are to be evaluated at the redshift of interest.  
We have already specified how, for a given halo mass, the virial 
radius depends on redshift; the NFW halo density profile is also 
specified by its concentration, which we assume is 
$c = 9\, (M/M_{*0})^{-0.13}/(1+z)$ \citep{2001MNRAS.321..559B}.
All this, in the right hand side of equation~(\ref{eq:wprp}), gives 
the halo model calculation of $w_p(r_p)$.  

Our halo model calculation of $\xi(s)$ makes two additional 
assumptions:  first, that satellite galaxies within haloes have isotropic 
velocity dispersions which are proportional to $GM/r_{\rm vir}$, and 
second, that the motion of the centre of mass of a halo is well 
described by linear theory.

\subsection{HOD fits}

\begin{table*}
  \begin{center}
    \caption{\label{tab:fithod8} The best fitting HODs to $w_p(r_p)$ assuming $\sigma_8$ = 0.8}
    \begin{tabular}{c  c  c  c  c  c  c  c  c  c} 
      \multicolumn{1}{c}{Selection} &
      \multicolumn{1}{c}{Redshift} &
      \multicolumn{1}{c}{Density} &
      \multicolumn{1}{c}{M$_{min}$} &
      \multicolumn{1}{c}{M$_1$} &
      \multicolumn{1}{c}{$\alpha$} &
      \multicolumn{1}{c}{$\chi^2_{red}$}&
      \multicolumn{1}{c}{$b_{lin}$}&
      \multicolumn{1}{c}{$M_{eff}$}&
      \multicolumn{1}{c}{$F_{sat}$}\\
      \multicolumn{1}{c}{} &
      \multicolumn{1}{c}{} &
      \multicolumn{1}{c}{($10^{-4}h^{3}$Mpc$^{-3}$)} &
      \multicolumn{1}{c}{($10^{13}M_{\sun}$)} &
      \multicolumn{1}{c}{($10^{13}M_{\sun}$)} &
      \multicolumn{1}{c}{} &
      \multicolumn{1}{c}{}&
      \multicolumn{1}{c}{}&
      \multicolumn{1}{c}{($10^{13}M_{\sun}$)}&
      \multicolumn{1}{c}{(\%)}\\
      \hline \hline 
	SDSS  & 0.21 & 0.94 $\pm$ 0.01 & 3.80 $\pm$ 0.07 & 34.2 $\pm$ 2.1 & 1.67 $\pm$ 0.23 & 1.22 & 2.11 $\pm$ 0.03 & 9.52 $\pm$ 0.59& 10.1 $\pm$ 3.7\\
	SDSS  & 0.55 & 0.73 $\pm$ 0.02 & 3.46 $\pm$ 0.06 & 34.0 $\pm$ 2.5 & 2.10 $\pm$ 0.38 & 1.12 & 2.42 $\pm$ 0.05 & 6.24 $\pm$ 0.51 & 4.7 $\pm$ 2.5\\
	2SLAQ & 0.19 & 1.64 $\pm$ 0.01 & 2.44 $\pm$ 0.02 & 27.0 $\pm$ 1.1 & 1.58 $\pm$ 0.13 & 0.77 & 1.91 $\pm$ 0.02 & 7.62 $\pm$ 0.41 & 10.4 $\pm$ 2.1\\
	2SLAQ & 0.55 & 1.65 $\pm$ 0.03 & 1.88 $\pm$ 0.02 & 21.8 $\pm$ 1.5 & 2.02 $\pm$ 0.2 & 1.23 & 2.16 $\pm$ 0.03 & 4.76 $\pm$ 0.20 & 6.2 $\pm$ 2.3\\
      \hline
    \end{tabular}
  \end{center}
\end{table*}

\begin{table*}
  \begin{center}
    \caption{\label{tab:fithod9}  The best fitting HODs to $w_p(r_p)$ assuming $\sigma_8$ = 0.9}
    \begin{tabular}{c  c  c  c  c  c  c  c  c  c} 
      \multicolumn{1}{c}{Selection} &
      \multicolumn{1}{c}{Redshift} &
      \multicolumn{1}{c}{Density} &
      \multicolumn{1}{c}{M$_{min}$} &
      \multicolumn{1}{c}{M$_1$} &
      \multicolumn{1}{c}{$\alpha$} &
      \multicolumn{1}{c}{$\chi^2_{red}$}&
      \multicolumn{1}{c}{$b_{lin}$}&
      \multicolumn{1}{c}{$M_{eff}$}&
      \multicolumn{1}{c}{$F_{sat}$}\\
      \multicolumn{1}{c}{} &
      \multicolumn{1}{c}{} &
      \multicolumn{1}{c}{($10^{-4}h^{3}$Mpc$^{-3}$)} &
      \multicolumn{1}{c}{($10^{13}M_{\sun}$)} &
      \multicolumn{1}{c}{($10^{13}M_{\sun}$)} &
      \multicolumn{1}{c}{} &
      \multicolumn{1}{c}{}&
      \multicolumn{1}{c}{}&
      \multicolumn{1}{c}{($10^{13}M_{\sun}$)}&
      \multicolumn{1}{c}{(\%)}\\
      \hline \hline 
	SDSS  & 0.21 & 0.94 $\pm$ 0.01 & 4.43 $\pm$ 0.15 & 45.5 $\pm$ 5.7 & 1.38 $\pm$ 0.16 & 1.45 & 1.91 $\pm$ 0.03 & 11.82 $\pm$ 0.60& 11.8 $\pm$ 2.2\\
	SDSS  & 0.55 & 0.73 $\pm$ 0.02 & 4.15 $\pm$ 0.09 & 46.3 $\pm$ 3.9 & 1.91 $\pm$ 0.39 & 1.13 & 2.20 $\pm$ 0.04 & 8.22 $\pm$ 0.81 & 5.1 $\pm$ 2.8\\
	2SLAQ & 0.19 & 1.64 $\pm$ 0.01 & 2.77 $\pm$ 0.03 & 34.2 $\pm$ 1.3 & 1.38 $\pm$ 0.13 & 0.97 & 1.73 $\pm$ 0.02 & 9.59 $\pm$ 0.68 & 11.7 $\pm$ 2.3\\
	2SLAQ & 0.55 & 1.65 $\pm$ 0.03 & 2.19 $\pm$ 0.03 & 28.2 $\pm$ 2.1 & 1.86 $\pm$ 0.20 & 1.26 & 1.96 $\pm$ 0.02 & 6.28 $\pm$ 0.36 & 6.8 $\pm$ 2.4\\
      \hline
    \end{tabular}
  \end{center}
\end{table*}

We fit for the parameters $M_{\rm min}$, $M_1$ and $\alpha$ (see equations~\ref{eq:Ncen} and~\ref{eq:Nsat}) by minimising a $\chi^2$ defined as the sum of the squared difference between the predicted and measured $n_g$ and $w(r_p)$ for a range of $r_p$. We use $w(r_p)$ rather than $\xi(r)$ as the numerical inversion required to calculate $\xi(r)$ increases the uncertainties and systematically reduces the slope in our power law fits.  Our fitting makes use of the full covariance matrices over 0.32 $< r_p <$ 50 $h^{-1}$ Mpc. We exclude scales smaller than 0.32 $h^{-1}$ Mpc as we are not confident that we have sufficiently corrected for fibre collisions. We note that the best fitting parameters are not significantly changed if the smallest bin included in the fit is one smaller or larger. 

The errors on the fits are determined by finding the region of parameter space with a $\delta \chi^2 \leq$ 1 (1$\sigma$ for 1 degree of freedom) from the best fit and then determining the maximum and minimum parameter values within that region. For $b_{lin}$, $M_{eff}$, and $F_{sat}$, which depend on all three of the fit parameters, the region used contains $\delta \chi^2 \leq$ 3.53 (1$\sigma$ for 3 degrees of freedom).

The resulting best fits are shown in Figure \ref{fig:wphod} and the best fit values for the HOD parameters are given in Table \ref{tab:fithod8}. We have checked that our best fitting model also provides a good description of our measurements of $\xi(s)$ and $\xi(r)$.  These parameters were not included in our definition of $\chi^2$ because the halo model of $\xi(s)$ requires further assumptions than does $w(r_p)$.  
Table~\ref{tab:fithod8} also provides the associated values of $F_{\rm sat}$, 
$M_{\rm eff}$ and $b_{\rm lin}$.

\begin{figure*}
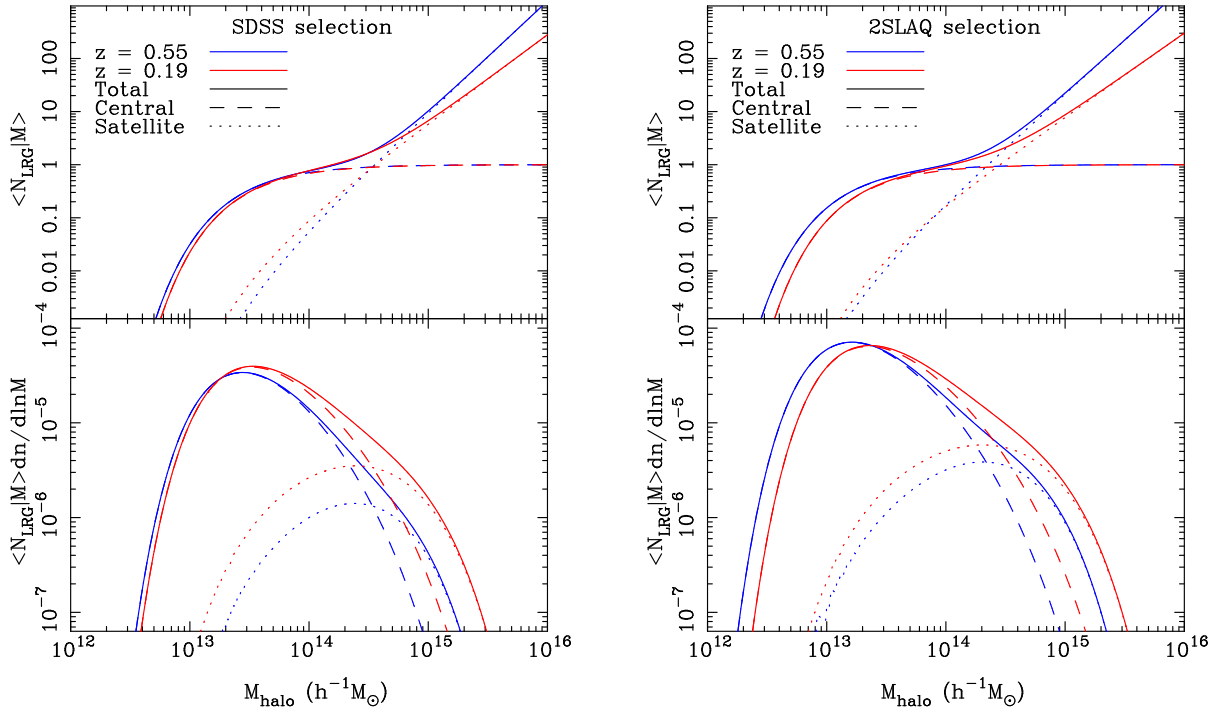

\vspace{10.5cm}
\includegraphics{NM_sdss_paper.ps}
\includegraphics{NM_2slaq_paper.ps}

\caption{\label{fig:NMhod} The mean number of LRGs per halo as a function of halo mass (top)  and the mean number of LRGs per halo times the number density of haloes as a function of mass (bottom) at $z\sim0.2$ (red) and $z\sim0.55$ (blue) for the SDSS selection matched (left) and for the 2SLAQ selection matched (right) samples. The total, central and satellite contributions are shown by the solid, dashed and dotted lines respectively.}
\end{figure*}

The best fitting HODs are shown in Figure \ref{fig:NMhod}.  
Increasing $\sigma_8$ (see Table \ref{tab:fithod9}) increases $M_{min}$ and $M_1$, and decreases $\alpha$. The bias decreases to compensate for the increased clustering strength of the dark matter, and $M_{eff}$ increases because $M_{min}$ is larger. The satellite fraction remains approximately the same, as $\alpha$ has reduced to compensate for the increase in $M_1$.

For our standard choice of $\sigma_8 =$ 0.8, the LRGs populate haloes with masses of order $10^{13} - 10^{14} M_{\odot}$; most of these LRGs are central galaxies - the satellite fractions are typically less than 10\%.  In the lower redshift samples $M_{eff}$ is larger by about 50\%, the bias is smaller by about 10\%, and the satellite fraction has approximately doubled.  The growth in $M_{eff}$ is a consequence of a 10\% increase in $M_{min}$, a small decrease in $M_1/M_{min}$, and a significant decrease in $\alpha$.  

It might seem paradoxical that decreasing $\alpha$ increases the satellite fraction.  This is a consequence of the fact that $M_1$ is larger than the mass scale on which the halo mass function drops exponentially (for $\sigma_8=0.8$, this scale is $0.6\times 10^{12}h^{-1}M_\odot$ and $1.9\times 10^{12}h^{-1}M_\odot$ at $z=0.55$ and $z=0.2$ respectively; when $\sigma_8=0.9$, these masses become $1.3\times 10^{12}h^{-1}M_\odot$ and $3.9\times 10^{12}h^{-1}M_\odot$).  Thus, increasing $\alpha$ increases the number of satellites in (the exponentially rare) haloes more massive than $M_1$ but {\em decreases} the number in less massive haloes which are exponentially more abundant.  

The larger satellite fractions at low redshift are best understood by thinking of the central and satellite populations separately.  If there is no merging, then the high $z$ satellites are satellites even at low $z$, whereas some of the high $z$ centrals have become satellites at low $z$ (e.g., if their host halo merged with a more massive halo).  As a result, the satellite fraction increases.  Merging would act in the opposite sense (satellites merging with satellites or with centrals would both reduce the satellite fraction).  

We note that the best fitting HODs for the $z=0.55$ samples are in excellent agreement with those presented in \citet{2008MNRAS.tmp..345B} who fit HODs to the angular clustering of 380,000 LRGs selected using the 2SLAQ LRG selection criteria with photometric redshifts $0.45 < z_{phot} < 0.65$. 

\section{Constraining LRG mergers}
\label{sec:HODmerge}

Paper I demonstrated that the evolution of the LF of LRGs was consistent with passive evolution of the stellar populations, and did not require any merging. If true, then as discussed in Section \ref{sec:nomerge}, the bias should evolve as $b(z_{lo}) = 1+ (b(z_{hi})-1) D(z_{hi})/D(z_{lo})$ where $D(z)$ is the growth factor \citep{1996MNRAS.282..347M,1996ApJ...461L..65F}. 
When applied to the bias of the best fitting $z=0.55$ HODs for the two samples, the predicted bias factors are 1.98 $\pm$ 0.02 at $z=0.19$ for the 2SLAQ selected sample and 2.20 $\pm$ 0.04 at $z=0.21$ for the SDSS selected sample. Both these values are significantly larger than the measured values given in Table \ref{tab:fithod8}, with the evolution in the 2SLAQ selected sample bias being incompatible with no-merging hypothesis at a significance of 98.4\%.  This is at a higher significance level to that calculated in Section \ref{sec:nomerge} using just the ratio of the large-scale clustering; the inclusion of the number density constraints in the HOD fits results in significantly smaller relative errors on the bias measurements than would be derived using clustering alone.

This argument against pure passive evolution still uses only the large-scale clustering signals at the two epochs.  In what follows, we use the language of the halo model to show that the evolution of the small-scale clustering signal also contains interesting information, and can provide even greater constraints on the importance of merging.

\subsection{HOD Evolution: No mergers}
\label{nomergers}
If we specify how galaxies populate haloes at some early time, $\langle N|m\rangle$, then we can estimate how this evolves as the haloes merge.  If the haloes merge but the galaxies do not, then 
\begin{equation}
 \langle N|M\rangle = \int_0^M {\rm d}m\,N(m|M)\,\langle N|m\rangle 
                    = C(M) + S(M) 
 \label{eq:C+S}
\end{equation}
where $N(m|M)$ is the mean number of haloes of mass $m$ which are in haloes of mass $M$ at the later time, and 
\begin{eqnarray}
 C(M) &=& \int_0^M {\rm d}m\,N(m|M)\,\langle N_c|m\rangle \qquad {\rm and}\\
 S(M) &=& \int_0^M {\rm d}m\,N(m|M)\,\langle N_c|m\rangle\,\langle N_s|m\rangle;
\end{eqnarray}
For $N(m|M)$ we use the expressions given by Sheth \& Tormen (2002), which generalise those of Lacey \& Cole (1993).  The Appendix shows that this guarantees that the comoving density $n_g$ is constant, whereas the large scale bias evolves in accordance with the continuity equation.  

\begin{figure}
\vspace{9.5cm}
\includegraphics{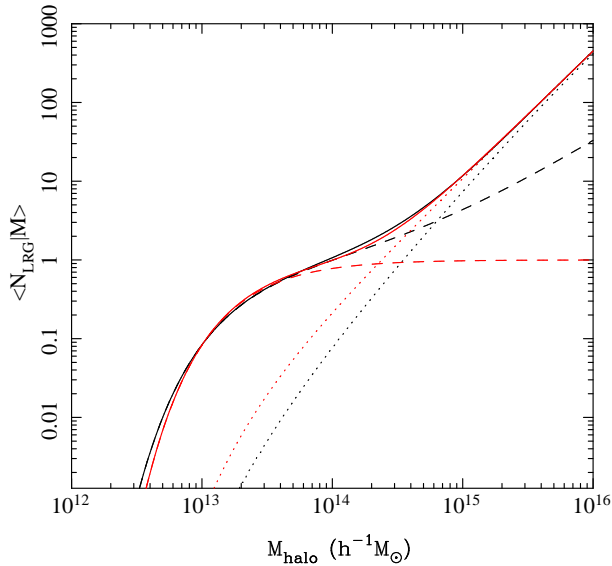}
\vspace{-1cm}
\caption{\label{fig:nomergers} The mean number of LRGs as a function of halo mass at $z\sim 0.19$ obtained by passively evolving the best fitting $z=0.55$ 2SLAQ HOD to $z=0.19$.  Upper dashed and lower dotted lines show the contributions from objects which used to be centrals and satellites; they sum to give the solid curve which  drops to zero at smaller mass scales; lower dashed and upper dotted lines, which sum to give the other solid curve, show the result of fitting this $\langle N|M\rangle$ to the form given in equations~(\ref{eq:Ncen}) and~(\ref{eq:Nsat}).}
\end{figure}

Whereas $C(M)$ counts the objects which used to be centrals, $S(M)$ counts the satellites.  Note that although $\langle N_c|m\rangle\le 1$, there is no guarantee that $C(M)\le 1$; indeed, for $M\gg M_{min}$, one expects $C(M)\ge 1$.  Figure~\ref{fig:nomergers} shows this explicitly; at late times, massive haloes may host many galaxies which were centrals at the earlier time.  

If we force $\langle N|M\rangle$ to have the same functional form as $\langle N|m\rangle$, then we can fit for $M_{min}$, $M_1$ and $\alpha$ at the later time.  These fitted values can then be inserted into the halo model calculation of $\xi$.  The Figure \ref{fig:nomergers} shows that forcing this parametrisation allows a good but not perfect description of the passively evolved HOD: the passively evolved HOD has 
a more gradual transition from 0 to 1.

It will turn out that, for the present study, it is important to accurately model this transition.  This is because we are studying rare objects which populate the high-mass end of the mass function.  As a result, haloes which host zero or one galaxies are substantially more numerous than those which host more.  Hence, allowing some lower mass haloes to host more than one galaxy (while making more such haloes void of galaxies) can affect the number of small separation pairs substantially.  

To illustrate this effect, let $p_0(M)$ denote the probability that 
a halo of mass $M$ contains no galaxies which were centrals at the 
higher redshift.  Then 
\begin{eqnarray}
 \label{eq:NcM}
 \langle N_c|M\rangle &=& 1 - p_0(M) \qquad{\rm and}\\
 \langle N_c|M\rangle\langle N_s|M\rangle &=&
  S(M) + C(M) - \langle N_c|M\rangle.
\end{eqnarray} 
The second equation assumes that only one of the high-$z$ centrals in a 
halo continues to count as the low-$z$ central; the others (of which 
there are $C(M) - \langle N_c|M\rangle$ on average) count as low-$z$ 
satellites.  
The mean galaxy count $\langle N|M\rangle$ is given by inserting 
these expressions in equation~(\ref{eq:Ntot}).  
This exercise shows that the problem is to model $p_0(M)$; the next subsection 
studies three different models.  

\subsection{HOD Evolution:  Small scale clustering and the 
            abundance of empty haloes}\label{p0models}

The quantity $p_0(M)$ counts the number of haloes of mass $M$ which 
were formed from mergers of objects which contained no galaxies.  
If the threshold $M_{\rm min}$ were sharp, then this would be simply 
related to the number of haloes at low redshift which did not have a single 
high-redshift progenitor of mass greater than $M_{\rm min}$.  
\citet{1999MNRAS.304..767S} have studied this problem; they provide expressions 
for the $k$-th factorial moment $\mu_k$ of the progenitor distribution.  
(Results in \citet{2002MNRAS.333..730C} suggest that these expressions are quite 
accurate.)  In principle, these can be used to estimate $p_0$, since 
 $p_0 = 1 + \sum_k (-1)^k\,\mu_k/k!$, where the sum runs from $k=1$ 
to an upper limit which is set by mass conservation; a halo of mass 
$M$ can have at most $M/M_{\rm min}$ progenitors. If the HOD where a step function 
then $M_{\rm min}$ would be the same as in Equation \ref{eq:Ncen}, else, it need not be.   
In practise, this is a complicated sum, so we have studied a 
few simpler models.  

\begin{figure}
\vspace{10.5cm}
\includegraphics{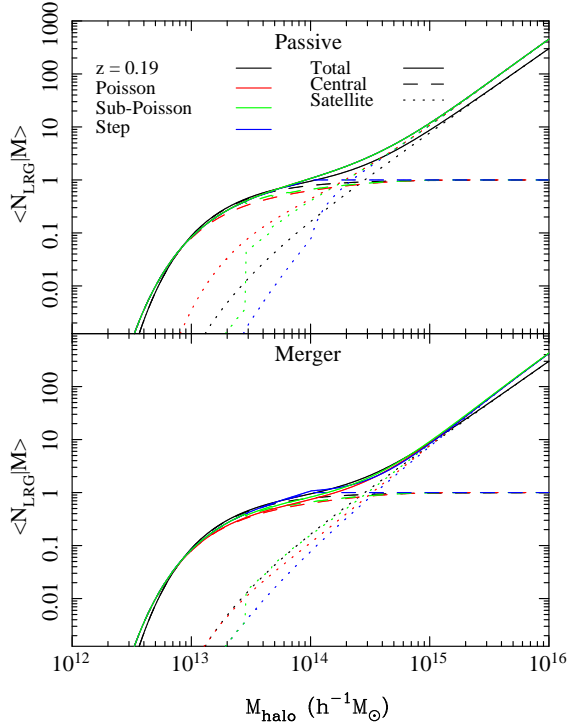}

\caption{\label{fig:halopM} The mean number of LRGs per halo as a function of halo mass at $z=0.19$ for the 2SLAQ selection matched samples. The top panel shows the effect of passively evolving the $z=0.55$ HOD to $z=0.19$ using the three models for $p_0(M)$ along with the measured HOD from the $z=0.19$. The effect of including merging of the central galaxies for the same models is shown in the bottom panel. The total, central and satellite contributions are shown by the solid, dashed and dotted lines, respectively.}
\end{figure}

\begin{figure}
\vspace{10.5cm}

\includegraphics{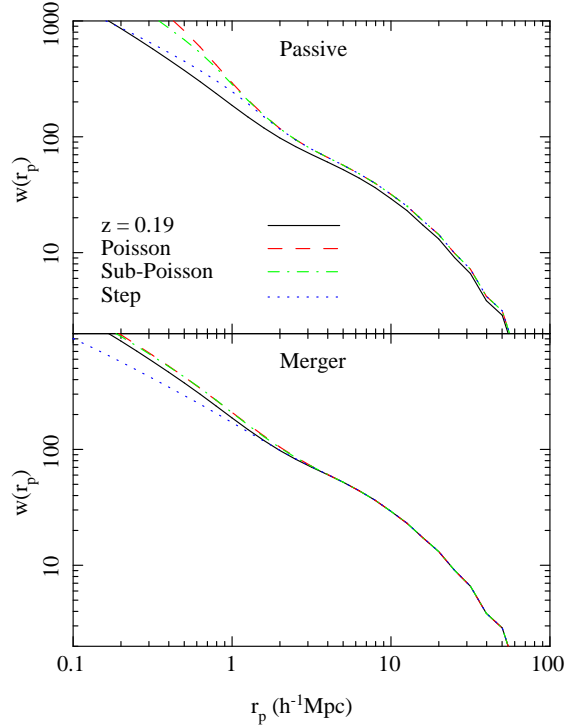}

\caption{\label{fig:2ptpM}The projected correlation function resulting from the evolving the $z=0.55$ 2SLAQ selection matched sample HOD to $z=0.19$ using the three models for $p_0(M)$ along with the correlation function from the HOD fit to the $z=0.19$ sample. The the effect of including merging of the central galaxies for the same models is shown in the bottom panel.}
\end{figure}

In our first model, we set 
\begin{equation}
 p_0(M) = {\rm e}^{-C(M)}.
\end{equation} 
This would be appropriate if the distribution of the number of 
high-redshift centrals in low-redshift haloes were Poisson (so $\mu_k=\mu_1^k$), 
with mean $\mu_1=C(M)$, and only one of these centrals continues to 
count as the low-redshift central; the others count as low-$z$ satellites.  
Note that if $C\ll 1$, then $\langle N_c|M\rangle\to C(M)$, so there 
is no correction to the satellite counts.  And if $C\gg 1$ then 
$\langle N_c|M\rangle\to 1$ and the satellite counts are increased 
by $C-1$.  Thus, our model interpolates smoothly between these 
two sensible limits. We show the resulting evolution in the way galaxies populate haloes and in the clustering in the top panels of Figures \ref{fig:halopM} and \ref{fig:2ptpM} as the red lines.

We have also studied what happens if, instead, we require a sharp 
transition between these two limits:  
set $\langle N_c|M\rangle = C(M)$ and $\langle N_s|M\rangle = S(M)$ 
when $C(M)\le 1$, and 
$\langle N_c|M\rangle = 1$ and $\langle N_s|M\rangle = S(M) + C(M)-1$ 
otherwise.  Compared to the Poisson model, this model has many more 
low-redshift haloes which host a single central high-redshift galaxy, and few which 
host more than one such galaxy; the Poisson model has fewer haloes 
which host galaxies, each allowed to host more than one high-redshift central.  
This decreases the number of high-redshift central pairs in haloes (compared 
to the Poisson model), which means that the number of central-satellite 
pairs is decreased, thus decreasing the small scale clustering signal.  
(Of course, higher-order statistics will also be affected:  
the probability of finding a large region devoid of galaxies will be 
larger in the Poisson model.) This model is plotted as the blue lines 
in the top panels of Figures \ref{fig:halopM} and \ref{fig:2ptpM}.

Whereas this second model is perhaps too simple, the Poisson model 
almost certainly allows too many low mass haloes to contain more than 
one galaxy, thus resulting in too many small scale pairs.  Indeed, 
mass conservation arguments \citep{1999MNRAS.304..767S,2002MNRAS.333..730C} 
strongly suggest that the progenitor counts should be sub-Poisson 
($\mu_k<\mu_1^k$), especially at low masses.  Furthermore, sub-Poisson 
counts are clearly seen in the numerical models 10 and 30 of \citet{2007arXiv0712.1643S}.  
The following Binomial model conserves mass, and lies between these 
two extremes:
\begin{equation}
 p_0(M) = \left[1-{C(M)\over N_{\rm max}}\right]^{N_{\rm max}} 
\end{equation}
where $N_{\rm max}= {\rm int}(M/M_{\rm min})$.  
We use this model as written for illustrative purposes only:  
in reality $M_{\rm min}$ is unlikely to be the same quantity as in 
equation~(\ref{eq:Ncen}), and the integer changes in $N_{\rm max}$ 
as $M$ increases produce artificial discontinuities in $\langle N|M\rangle$.  
Nevertheless, this model predicts a small scale clustering signal 
which lies below that associated with the Poisson model, but 
above that for the sharp threshold model shown as the green lines 
in the top panels of Figures \ref{fig:halopM} and \ref{fig:2ptpM}.  

The top panel of Figure \ref{fig:2ptpM} shows that for all three models for $p_0(M)$ the passive evolution of the clustering predicts a far greater increase in the clustering strength than is observed. This is caused by the presence of too many satellite galaxies, with satellite fractions of 27$\pm$3\%, 11$\pm$1\% and 19$\pm$2\% for the Poisson, Step and Sub-Poisson models respectively compared to 10$\pm$2\% for the best fitting HOD to the data.

\subsection{HOD Evolution: Central-central mergers}
Once we have decided how likely it is that a low-redshift halo contains at 
least one high-redshift central galaxy, we also study models in which centrals 
merge onto centrals.  
This is motivated by the fact that central galaxies are expected to 
be more massive than satellites, so dynamical friction may be more 
effective at making these objects merge onto the true low-redshift central.  
To model this case, we again use equation~(\ref{eq:NcM}) for 
$\langle N_c|M\rangle$, but we set 
\begin{equation}
 \langle N_c|M\rangle\langle N_s|M\rangle = S(M) + 
    f_{\rm no-merge}\,\Bigl[C(M)-\langle N_c|M\rangle\Bigr],
\end{equation} 
where $f_{\rm no-merge}$ is the fraction of low-redshift satellites 
which were high-redshift centrals, and have not merged with one another 
or onto the new central object.  

When $f_{\rm no-merge}=1$ then this 
is the same as the no merger model of the previous section; when 
$f_{\rm no-merge} = 0$, then the central galaxies of all the high-redshift 
haloes which merged to make a low-redshift halo have merged to make a single 
massive central galaxy.  Strictly speaking, the model says nothing 
about what these objects merged with - they may have merged with 
one another or with other satellites - it only assumes that the 
number of objects which merge scales with $M$ in the manner given 
above.  However, the assumption that they merged onto the central 
object has considerable physical appeal.

\begin{figure}
\vspace{10.5cm}
\includegraphics{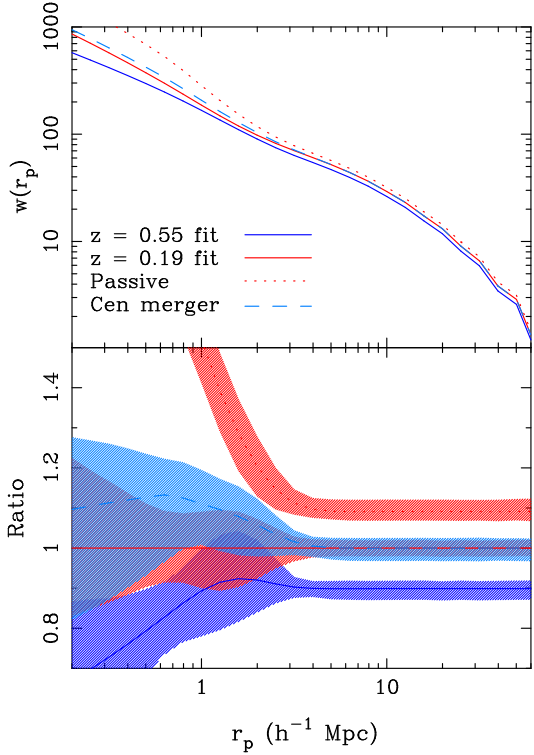}

\caption{\label{fig:hodevo} Halo model fits to w($r_p$) at $z\simeq0.19$ (red solid) and $z\simeq0.55$ (blue solid) for the 2SLAQ selection matched sample. The effect of passively evolving the $z=0.55$ HOD to $z=0.19$ is shown as the dotted line and the effect of including merging of the central galaxies is shown as the dashed line. The bottom panel shows the ratios of the w($r_p$) fits shown above to the measured {\it z} = 0.19 fit. The shaded areas enclose the 1 sigma confidence regions.}
\end{figure}

The result of applying this merger model for the three parametrisation of $p_0(M)$ are shown in the bottom panels of Figures \ref{fig:halopM} and \ref{fig:2ptpM}. For each model we chose the value of $f_{no-merge}$ that best matches the large scale clustering, 0.1, 0, 0.25 for the Poisson, Step and Sub-Poisson models respectively. In all cases the agreement in the high mass haloes is much improved and the satellite fraction reduces to 11$\pm$3\%, 7$\pm$2\%, 10$\pm$3\%, comparable to the meaured value. The best fit at small scales is provided by the sub-Poisson model; this is reasuring, as it is the most physically motivated - although our implementation is not yet ideal. This suggests that the data are consistent with a generic prediction of hierarchical models - that the scatter in merger histories should produce sub-Poisson scatter. The step model produces far too little small scale clustering, consistent with its lower satellite fraction, with both the Poisson and sub-Poisson models providing a reasonable match within the errors. 

We show in Figure \ref{fig:hodevo} a more detailed comparison of the passive and merger sub-Poisson model with the measured correlation functions by dividing each by the best fit to the {\it z} = 0.19 measurement. Also shown are the 1$\sigma$ confidence regions calculated by propagating the error on the fit at {\it z} = 0.55. This figure explicitly shows that the passive model is ruled out at high significance. On large scales ($>$ 3 $h^{-1}$) Mpc the passive model is incompatible with the measured clustering at $z$ = 0.2 at the 98\% level, consistent with the constraints from the bias evolution given above. However, when smaller scales are included the passive model becomes increasingly incompatible with the measured clustering; for scales larger than  1 $h^{-1}$ Mpc the passive model is excluded at a confidence level of greater than 99.9\%, with the level of significance increasing with the inclusion of even smaller scales. The sub-Poisson merger model is consistent with the data on all scales, even though the fraction of centrals which are alowed to merge is determined by matching only the large-scale clustering.

We have demonstrated that it is necessary to allow some merging (or some other method of removal) of some fraction of the high redshift LRGs if we wish to reproduce the clustering at low redshift. This will have the effect of reducing the space density of the evolved population at low-redshift, something that we do not observe in the data. The change in the space density associated with the best fitting sub-Poisson model is 9.2$\pm$2.6\%, suggesting that at most about 20\% of the LRGs are merging with each other. In fact there are on average 2.34 high-redshift centrals in each merged halo, resulting in 16.1$\pm$4.6\% of LRGs experiencing an LRG-LRG merger. This is consistent with the constrainst provided by the luminosity function evolution of Paper I. For comparison the Possion model predicts a change in the space density of 19.4$\pm$5.5\%, suggesting that up to 40\% of LRGs have been involved in a LRG-LRG merger. This number is highly inconsistent with the LF measurments and lends further support to the sub-Poisson model.

If we continue with the hypothesis that LRGs are merging with one another, it is reasonable to assume that some red galaxies too faint to be included in our sample at {\it z} = 0.55 will have also merged by {\it z} = 0.19, some of which will now be sufficiently luminous to be included in that sample. These galaxies will then increase the space density of the low-redshift LRG sample, potentially allowing the space density to remain unchanged. From the measurements we have no constraints on how many of these galaxies there are and how they are distributed within the dark matter haloes and thus how they might change the clustering. Because the space density has changed in the merger model one could argue that we should compare our evolved high redshift 2pt correlation function with one measured from a sample of low redshift LRGs with a matching lower space density. The difficulty with this approach is deciding which galaxies to remove from our observed sample in order to reduce the space density. 

An obvious choice would be to change the magnitude limit, thus removing the galaxies with the lowest stellar masses, equivalent to the approach taken in \citet{2007ApJ...655L..69W}. However, in our merging model, we merge high redshift central galaxies, and it seems unlikely that these would represent the LRGs with the lowest stellar masses. Alternatively, if we randomly sample the low redshift HOD, we will reduce the space density with out changing the clustering. This is equivalent to saying that the LRGs, which are newly formed by the merging of lower luminosity red galaxies at low redshift, trace the dark matter in the same way as the whole LRG population. If this is a true reflection of the evolution of the LRG population, then the randomly sampled measured HOD should look like the HOD produced by our central merging model. 

\begin{figure*}
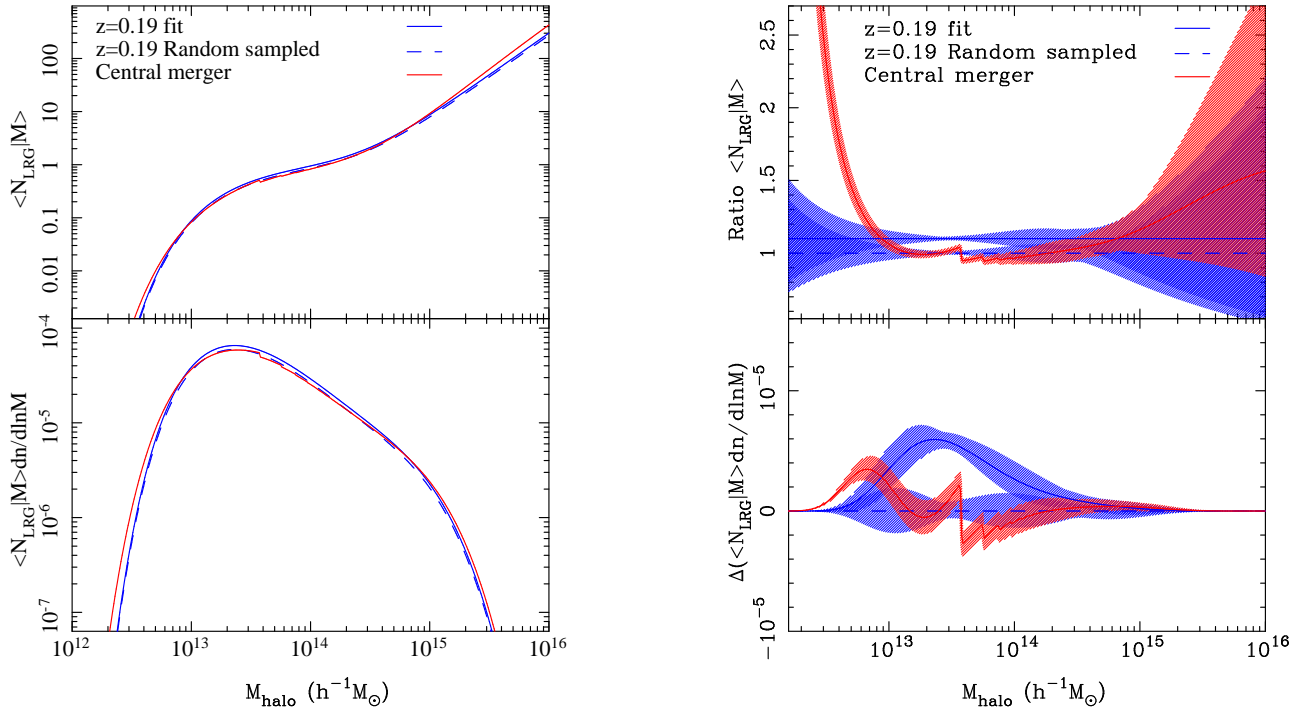

\vspace{10.5cm}

\includegraphics{NM_randomsample.ps}
\includegraphics{NM_randomsample_ratio.ps}

\caption{\label{fig:halorand}In each panel the blue solid line shows the best fit to the 2SLAQ selection matched sample at {\it z} = 0.19. The effect of applying the sub-Poisson central merger model to the {\it z} = 0.55 HOD is shown as the red solid line and the {\it z} = 0.19 HOD fit random sampled to match the space density of the merger model is shown as the dashed blue line. The mean number of LRGs per halo as a function of halo mass is shown top left and the mean number of LRGs per halo times the number density of haloes as a function of mass is shown bottom left. The ratio of the numbers of LRGs is shown upper right and the difference in the space densities shown bottom right.}
\end{figure*}

We show in Figure \ref{fig:halorand} a comparison of the HOD of the best fit sub-Poisson merger model with the best HOD fit to the {\it z} = 0.19 measurement, along with the measured {\it z} = 0.19 HOD randomly sampled to match the space density of the merger model HOD. The left side of Figure \ref{fig:halorand} shows the HODs and the HODs weighted by the number density of the haloes in the same way as we've shown before. The right side shows the ratio of the HODs (top) and the difference between the weighted HODs (bottom). For all but the lowest masses there is reasonable agreement between the randomly sampled HOD and the merger HOD. At the low mass end, the large difference is due in part to our having to force the {\it z} = 0.19 HOD to have a particular functional form; a form which the central merger model is not required to satisfy. There is still some discrepancy beyond that caused by the steps introduced by the binomial form of the sub-Poisson model, suggesting that any newly formed LRGs, which have been added to the low redshift sample, do not trace the dark matter in exactly the same way as the existing LRGs.  

\section{Comparison with previous work}
\label{sec:comp}

\subsection{Merger Rates}

A number of authors have recently tried to constrain the merger rate of LRGs using a variety of methods.
\citet{2006ApJ...652..270B} estimate that  50\% of massive galaxies ($>5\times10^{10} M_{\sun}$) have experienced a major merger since {\it z} = 0.8. They also show that the merger rate increases with redshift and provide a fitting formula for this increase. Applying this formula to the redshift interval we are considering here yields a merger rate of 21\% between z=0.55 and z=0.19. The merger rate defined by \citet{2006ApJ...652..270B} is the equivalent of the change in space density we measure i.e. 9.2\%. However, \citet{2006ApJ...652..270B} sample has a space density of 33$\times10^{-4} h^{-1}$Mpc$^3$ which is 20 times higher than ours and thus consists of galaxies with typically much less stellar mass. The merger rate is believed to increase with decreasing stellar mass so any direct comparisons between the two measurements are difficult. 

\citet{2006ApJ...644...54M} use the small scale clustering to estimate an LRG-LRG merger rate of 0.625\% Gyr$^{-1}$ for SDSS LRGs at {\it z} = 0.25. This would correspond to 2\% from {\it z} = 0.55 to {\it z} = 0.19 far lower than our measurement. Applying the fitting formula for the evolution of the merger rate from \citet{2006ApJ...652..270B} normalised to match the \citet{2006ApJ...644...54M} value at {\it z} = 0.25 yields a rate $\simeq$4\%, still a factor of 2.5 lower than our best fitting value. Once again the galaxy samples aren't directly comparable since the space density of LRGs in the \citet{2006ApJ...644...54M} sample are a factor of 3.5 lower than the sample we use here, so one would expect the merger rate to be lower for the more massive \citet{2006ApJ...644...54M} LRGs. 

\citet{2007MNRAS.379.1491C} use N-body simulations to follow the accretion
of halos sufficiently massive to host LRGs. They then compare this
accretion history with the observed multiplicity function of LRGs at
$z\sim0.3$ \citep{2007arXiv0706.0727H} in order to constrain the LRG
merger timescale and hence merger rate. Using this method they find a
LRG-LRG merger rate approximately a factor of two higher than that
measured by \citet{2006ApJ...644...54M} using the small scale clustering
of LRGs.

Using a very similar methodology to our own, \citet{2007ApJ...655L..69W} estimate that $\sim$ 1/3 of the low-redshift satellite galaxies must be destroyed (e.g. merge) in order to match the clustering evolution of luminous red galaxies between {\it z} = 0.9 and {\it z} = 0.5 in the NDWFS. This corresponds to a merger rate of 3.4\% Gyr$^{-1}$, which would be 10.6\% over our redshift interval. This rate is comprable to our estimates; however, based on the \citet{2006ApJ...652..270B} trend, one would expect a factor of 2 increase in the mean rate due to the higher redshift of the \citet{2007ApJ...655L..69W} sample and also an increase due to the factor of 6 higher space density of their LRGs. There is, however, one important difference between the \citet{2007ApJ...655L..69W} study and the one presented here that may rectify some of the descrepency in the merger rates. As mentioned above \citet{2007ApJ...655L..69W} adjust the space density of the low redshift LRG sample HOD fit with which they compare to their evolved high redshift sample. This is accomplished by adjusting the mass scale of the HOD fit by 7\% to higher masses. This approach, of course, would reduce the space density and increase the clustering, resulting in a lower amount of merging required to reduce the clustering produced by the passive evolution model to the measured level. Reducing the fraction of high-redshift centrals allowed to merge in the model similarly increases the clustering but also decreases the space density. 

\begin{figure}
\vspace{7.5cm}

\includegraphics{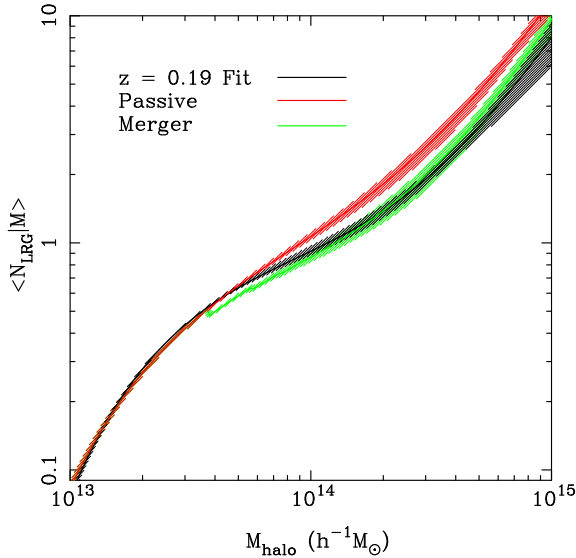}

\caption{\label{fig:NMms} The mean number of LRGs per halo as a function of halo mass at {\it z} = 0.19 (black) for the 2SLAQ selection matched sample with the mass scale increased by 6.4\%. The effect of passively evolving the {\it z} = 0.55 fit to {\it z} = 0.19 is shown as the red line and the effect of including merging of the central galaxies is shown as the green line.}
\end{figure}

\begin{figure}
\vspace{7.5cm}

\includegraphics{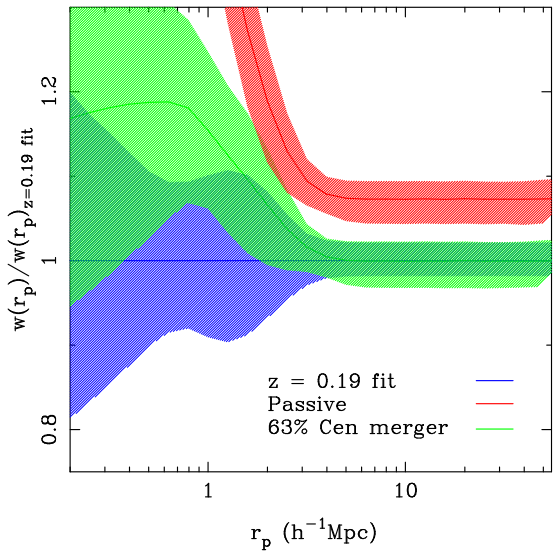}

\caption{\label{fig:Xims} The ratio of the projected correlation functions to the best fit HOD at {\it z} = 0.19 where the mass scale of the HOD has been increased by 6.4\%. The effect of passively evolving the {\it z} = 0.55 fit to {\it z} = 0.19 is shown as the red line and the effect of including merging of the central galaxies is shown as the green line. }
\end{figure}

Therefore there is only one unique combination of mass scale shift and merger rate that will match both the clustering and space density simultaneously. We find that increasing the {\it z} = 0.19 HOD mass scale by 6\% and allowing 63\% of the high-redshift centrals to merge yields a large-scale bias of 1.93 and space density of 1.52$\times10^{-4} h^{-1}$Mpc$^3$ for both the measured low-redshift HOD and the evolved high-redshift HOD. This corresponds to a merger rate of 7.5$\pm$2.3\% between {\it z} = 0.55 and {\it z} = 0.19. Figures \ref{fig:NMms} and \ref{fig:Xims} show the HOD and clustering respectively. Within the errors the merger model yields a good match with the measured HOD although the small scale clustering is a slightly poorer fit than the model with more merging shown in Figure \ref{fig:hodevo}. This value is now in better agreement to that which one might derive from the measurement of \citet{2006ApJ...644...54M} and the estimate of \citet{2007ApJ...655L..69W} although it still seems marginally higher. This may of course be due to the uncertainty in the dependence of the merger rate with redshift and mass. Alternatively the possible discrepancy with the \citet{2007ApJ...655L..69W} result, which uses a very similar method coupled to N-Body simulations could point to a deficiency in the current theoretical models of the conditional mass function used herein.\\

\begin{figure}
\vspace{7.0cm}

\includegraphics{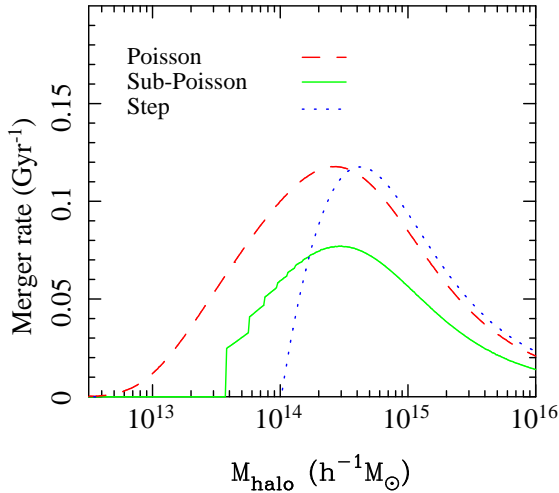}

\caption{\label{fig:Mfrac}The LRG merger rate as a function of halo mass.}
\end{figure}

Finally, \citet{2007arXiv0710.2157M} search for evidence of disturbance in close pairs of massive galaxies in $z <$ 0.12 groups to estimate the merger rate. They find that most of the mergers are occurring between approximately equal mass red progenitors and typically involve the central group galaxy, a picture that is consistent with the model we present here. They determine a merger rate 2-9 times higher than that of \citet{2006ApJ...644...54M} for comparable galaxies and suggest that this is because their minimum group mass is $3.5\times10^{13}M_{\sun}$, higher than the typical halo mass of LRGs, and therefore the merger rate of LRGs increases with increasing halo mass. We show in Figure \ref{fig:Mfrac} the merger rate as a function of halo mass for our three merger models. This figure does indeed indicate a rapid increase in the merger rate in haloes with mass up to 3 or 4$\times10^{13}M_{\sun}$, but with a decrease at higher masses.

\subsection{Semi-analytic Models}

\begin{figure}
\vspace{9.5cm}

\includegraphics{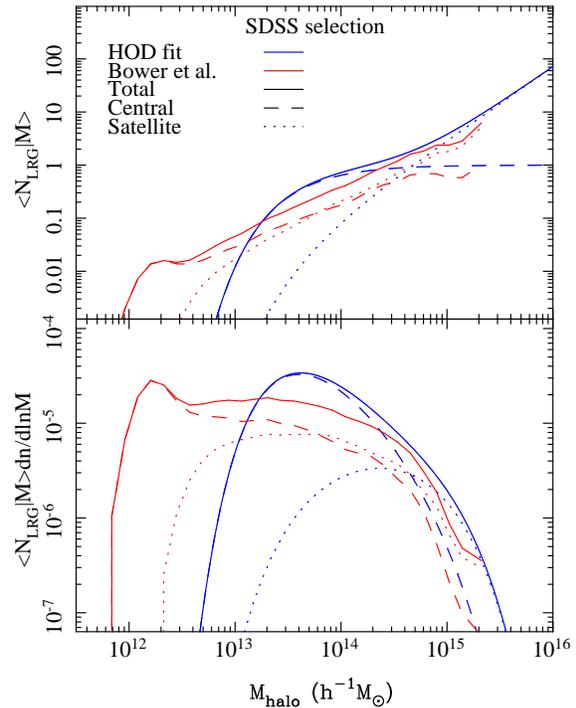}

\caption{\label{fig:NMmodel} The mean number of LRGs per halo as a function of halo mass (top) and the mean number of LRGs per halo times the number density of haloes as a function of mass (bottom) for the Bower et al. model of LRGs presented in \citet{2007arXiv0710.3557A} (red lines) compared to those generated from the fits to correlation functions present herein (blue lines). The total, central and satellite contributions are shown by the solid, dashed and dotted lines respectively.}
\end{figure}

\citet{2007arXiv0710.3557A} present a comparison of semi-analytic galaxy formation models to various properties of samples of LRGs very similar to the ones presented here. They find that the one of their models \citep{2006MNRAS.370..645B} gives a good match to the luminosity function of SDSS LRGs at {\it z} = 0.24, but over predicts the abundance of 2SLAQ LRGs at {\it z} = 0.55. The \citet{2006MNRAS.370..645B} model is also able to reproduce the clustering of samples at both {\it z} = 0.5 and at {\it z} = 0.24. They also present HODs generated from their models and compare them to the best fitting HODs for the samples presented within this paper, where the HOD fits are made using the same cosmological parameters as are used in the semi-analytic models. We only consider here the \citet{2006MNRAS.370..645B} z=0.24 HOD, which is shown in Figure \ref{fig:NMmodel} since it matches both the LF and clustering of SDSS selected LRGs at {\it z} = 0.24. The plotted HOD has a quite different form from the one we measure and is not reproducible with the formulation we have used in this paper.

In addition these models predict satellite fractions of 20-30\% which is a factor of 2-3 times higher than our HOD fits yield, but a merger rate for the 2SLAQ selected sample of $\sim$5\% over our redshift range, in good agreement with the observations. Plotting the central and satellite HODs separately for the \citet{2006MNRAS.370..645B} model (Figure \ref{fig:NMmodel}) demonstrates the reason for the high satellite fraction. There are many haloes that do not have an LRG central but do have satellite LRGs.  
It may appear surprising that the central galaxy within a halo does not
meet the LRG selection criteria. Although the central galaxy is the most
massive in terms of stellar mass and cold gas mass, it is not necessarily
the brightest in the observer frame r-band. 

A more likely scenario,
however, is the case in which the central galaxy is the brightest galaxy
in the halo, but does not match the LRG colour selection.  In the \citet{2006MNRAS.370..645B} model the suppression
of gas cooling by AGN heating ramps up gradually from intermediate mass
haloes, so some gas is still cooling in haloes with $M_{\rm halo} \sim
10^{12} h^{-1} M_{\odot}$ and being directed onto the central galaxy.
This supply of cold gas results in recent star formation in the central
galaxy. In more massive haloes, the cooling flow is suppressed more
strongly, so central galaxies in these haloes experience no recent star
formation.
 It may also be the case that there are too many red satellites due to the instantaneous stripping of the gas a galaxy experiences in the \citet{2006MNRAS.370..645B} model. However, one does need to remain cautious with these comparisons since even though at {\it z} = 0.24 the \citet{2006MNRAS.370..645B} model does match both the LF and clustering, it is unable to reproduce the evolution of the LF, suggesting that it is still lacking in some areas. Even so, it does suggest that the form of the HOD we are using may be too simplistic when a colour selection is included along with a luminosity cut. We will investigate this further in a forthcoming paper, which includes both a better treatment of the gas stripping \citep{Font2008} and a refined AGN feedback model.

\section{Summary and Conclusions}
\label{sec:discuss}

We present here a detailed analysis of the clustering of Luminous Red Galaxies (as defined by \citet{2001AJ....122.2267E} and \citet{2006MNRAS.372..425C}) as a function of redshift using samples of LRGs matched to have the same intrinsic colours and luminosities assuming passive evolution of their stellar populations. These galaxies represent the most massive in the universe with stellar masses lager than $10^{11} h^{-1} M_{\odot}$ and space densities of $\simeq 10^{-4} h^3 $Mpc$^{-3}$. We find that:  

\begin{itemize}
\item The amplitude of the clustering ($r_0$) does not significantly evolve with redshift over $0.15 < z < 0.6$, whereas there is a marginally significant decrease in the slope ($\gamma$) with increasing redshift.  \\
\item The lack of evolution in the clustering amplitude on large-scales is inconsistent with a picture in which the LRGs have purely passive evolution undergoing no major mergers over this time period, and rules out this passive model at $98\%$ significance.  \\
\item A HOD where the fraction of haloes which host central galaxies $\langle N_c|M\rangle = \exp(-M_{min}/M)$ and only haloes which host centrals can host satellites where the satellites are drawn from a Poisson distribution with mean $\langle N_s|M \rangle = (M/M_1)^{\alpha}$ is able to accurately reproduce the clustering and space density of our LRG samples. Within this framework the LRGs are predicted to be hosted in haloes with a typical mass close to $10^{14} h^{-1} M_{\odot}$ which increases by $\simeq$ 50\% from $z=0.55$ to $z=0.2$, and to have satellite fractions increasing from $\simeq$ 5 to 10\% over this time. The LRGs are found to have a bias $\simeq 2$ and which decreases with redshift at a much greater rate than would be predicted for the passive no merger case.\\
\item We introduce an analytic approach to describe the evolution of the HOD with redshift, and demonstrate that this guarantees that the comoving density remains constant and the large scale bias evolves in accordance with the continuity equation. We use this approach to further demonstrate that the passive evolution of the LRG HOD from {\it z} = 0.55 is inconsistent with the measurements at {\it z} = 0.19 at greater than 99.9\% significance, predicting far too many satellite galaxies at {\it z} = 0.19 and greatly over estimating the clustering strength on all scales.\\ 
\item We introduce a model in which high-redshift centrals are allowed to merge with other high-redshift centrals occupying the same halo at low-redshift. This choice is motivated by the fact that centrals are likely to be more massive than satellites, so dynamical friction may be more effective at making these objects merge with the true low-redshift central. This model is able to accurately match the large-scale clustering evolution of the LRGs. We demonstrate that the small scale clustering is dependent on the parametrisation of the scatter in halo merger histories. We investigate three models for this scatter and find that both the sub-Poisson and Poisson models are able to match the small-scale clustering evolution. However, the Poisson model requires a much larger LRG-LRG merger rate (20\%) which is not favoured by either the evolution of the LRG luminoisty function \citep{2006MNRAS.372..537W} or other independent measures of the LRG-LRG merger rate \citep{2006ApJ...644...54M,2007ApJ...655L..69W}. We therefore favour the best motivated sub-Poisson scatter giving observational support to this generic prediction of hierarchical models.\\
\item In order to match the clustering evolution we require an LRG-LRG merger rate of 7.5$\pm$2.3\% from {\it z} = 0.55 to {\it z} = 0.19 corresponding to 2.4\% Gyr$^{-1}$. This is probably consistent with other measurements of the merger rate of massive red galaxies given the uncertainties in how the merger rate depends on the mass of the galaxy and evolves with redshift.\\
\item Although some merging is required to match the clustering evolution, the merger rate is sufficiently small that it is entirely compatible with the low rate of evolution in the luminosity function of LRGs found in \citet{2006MNRAS.372..537W}.\\ 
\item We compare in detail the measured HOD for one of the LRG samples to that predicted by the latest semi-analytic models of galaxy formation for a very similar sample of LRGs which matches both the luminosity function and clustering as described in \citet{2007arXiv0710.3557A}. The model HOD is very different to our fit, and would not be reproducible by the functional form of the HOD we assume. In particular the model has many haloes that contain LRG satellites where the central is not an LRG. This suggests that a more sophisticated form of the HOD may be required for galaxy samples selected by colour in addition to luminosity, although caution is required as the semi-analytic model is still unable to accurately reproduce the evolution of the LRG population.\\
\item Our halo model analysis of the relation between the low- and 
high-redshift populations is similar in spirit to those of 
\citet{2007ApJ...655L..69W} and \citet{2007arXiv0712.1643S}.  
However, whereas their work used numerical simulations, our 
approach is entirely analytic.  This means that our analysis 
relies heavily on the accuracy of current models of $N(m|M)$, 
the conditional mass function.  These models are not particularly 
accurate for small redshift intervals \citet{2002MNRAS.329..61S}, 
so we hope that our analysis will generate interest in improving 
these models.  \\
\item Our analysis also highlights the need for a better 
understanding of the stochasticity in halo merger histories.\\
 
\end{itemize}

\section*{Acknowledgements}

The authors are very grateful to Shaun Cole, Britt Lundgren, Hee-jong Seo, Rob Smith, Russell Smith, and Martin White  for advice and comments on this work.  The authors thank the AAO staff for
their assistance during the collection of these data. We are also
grateful to the PPARC TAC and ATAC for their generous allocation of
telescope time to this project. This work was supported in part by a rolling grant from STFC and NSF grant 0507501. DAW thanks the Department of Astronomy at the University of Illinois for their regular hospitality. RKS thanks the Aspen Center for Physics for their hospitality. CMB is supported by the Royal Society. RCN was partially supported during this research by a EC Marie Curie Excellence Chair. 

We would like to thank Cameron McBride, Jeff Gardner, and Andy Connolly for providing a pre-release version of the Ntropy package.  Ntropy was funded by the NASA Advanced Information Systems Research Program grant NNG05GA60G.

Funding for the SDSS and SDSS-II has been provided by the Alfred P.
Sloan Foundation, the Participating Institutions, the National Science
Foundation, the U.S. Department of Energy, the National Aeronautics
and Space Administration, the Japanese Monbukagakusho, the Max Planck
Society, and the Higher Education Funding Council for England. The
SDSS Web Site is http://www.sdss.org/.

The SDSS is managed by the Astrophysical Research Consortium for the
Participating Institutions. The Participating Institutions are the
American Museum of Natural History, Astrophysical Institute Potsdam,
University of Basel, Cambridge University, Case Western Reserve
University, University of Chicago, Drexel University, Fermilab, the
Institute for Advanced Study, the Japan Participation Group, Johns
Hopkins University, the Joint Institute for Nuclear Astrophysics, the
Kavli Institute for Particle Astrophysics and Cosmology, the Korean
Scientist Group, the Chinese Academy of Sciences (LAMOST), Los Alamos
National Laboratory, the Max-Planck-Institute for Astronomy (MPIA), the
Max-Planck-Institute for Astrophysics (MPA), New Mexico State
University, Ohio State University, University of Pittsburgh,
University of Portsmouth, Princeton University, the United States
Naval Observatory, and the University of Washington.

\appendix
\section{Constant comoving number density in the halo model}
\subsection{Large scale clustering in real-space}
Let $g(m)$ denote the mean number of galaxies in haloes of mass
$m$ at some early time, and let $G(M)$ denote a similar quantity
at some later time.
If haloes merge but galaxies do not, then
\begin{equation}
  G(M) = \int_0^M {\rm d}m\,N(m|M)\, g(m)
\end{equation}
where $N(m|M)$ denotes the mean number of $m$ haloes from the
earlier epoch which have been incorporated into $M$ haloes by
the later epoch.
Lacey \& Cole (1993) and Sheth \& Tormen (2002) discuss
models for $N(m|M)$ that are consistent with the halo abundances
of Press \& Schechter (1974) and Sheth \& Tormen (1999).

To see that the number density of galaxies has indeed not changed,
note that
\begin{eqnarray}
  \bar n &\equiv& \int_0^\infty {\rm d}M\,n(M)\,G(M) \nonumber\\
   &=& \int_0^\infty {\rm d}M\,n(M)\int_0^M {\rm d}m\,N(m|M)\,g(m) 
\nonumber\\
   &=& \int_0^\infty {\rm d}m\,g(m)\int_m^\infty {\rm d}M\,n(M)\, 
N(m|M)\nonumber\\
   &=& \int_0^\infty {\rm d}m\,n(m)\,g(m).
\end{eqnarray}
The first equality expresses the number density as an integral over 
the low-redshift halo population, whereas the final equality integrates 
over the high-redshift population.  
The associated large scale bias factor at the later time is
\begin{eqnarray}
  b_0-1 &=& \int_0^\infty {\rm d}M\,{n(M)\,G(M)\over\bar n}\,[b(M)-1] 
\nonumber\\
   &=& \int_0^\infty {\rm d}m\,{g(m)\,n(m)\over\bar n}\nonumber\\
   && \quad\times \int_m^\infty {\rm d}M\,{n(M)\, N(m|M)\over 
n(m)}\,[b(M)-1].
\end{eqnarray}

Now,
\begin{equation}
  b(M) = 1 - {{\rm d}\ln n(M)\over {\rm d}\delta_c}
\end{equation}
(Sheth \& Tormen 1999) and the algebra in Abbas \& Sheth (2005)
shows that the expression above reduces to
\begin{eqnarray}
  b_0-1 &=& \int_0^\infty {\rm d}m\,{g(m)\,n(m)\over\bar n}\,
                          {[b(m)-1]\over D_0/D_z}
      \nonumber\\
        &=& (b_z-1)/(D_0/D_z)
\end{eqnarray}
where $D$ is the linear theory growth factor.
(If the later time is the present in an Einstein de-Sitter
universe, then $D_0/D_z = a_0/a_z = 1+z$.)
This shows explicitly that the halo model calculation of the
evolution of the bias in the no-merger model is the same as that
derived from an argument based on the continuity equation
(Nusser \& Davis 1994; Fry 1996).  Note that the bias factor
evolves even though the number density does not.

One might wonder if, although the bias factor evolves, the clustering
strength itself does not.  The ratio of the large-scale clustering
signal at the two epochs is
\begin{equation}
  {\xi_0(r)\over\xi_z(r)} = {b_0^2 D_0^2\over b_z^2 D_z^2}
  = \left({b_0\over b_0 - 1 + D_z/D_0}\right)^2;
\end{equation}
since $D_z<D_0$, the later epoch is more strongly clustered.  
For example, for $b_0 = 2$ and $D_z/D_0 = 2/3$, this factor is $(6/5)^2 = 1.44$.
Setting $D_z/D_0\ll 1$ illustrates a fact that is often overlooked:
the clustering strength of highly biased objects (i.e., the most
massive haloes) evolves very little, even though the clustering of
the dark matter itself has evolved significantly: $(D_0/D_z)^2 \gg 1$.
The most massive objects do not move far from their initial comoving
positions.

This calculation suggests a simple test of the null hypothesis that
two populations having the same comoving number density are related
by the no-merger evolution model:  If the measured clustering
signal has not evolved, or if the high redshift sample is more 
strongly clustered, then the hypothesis can be rejected.

\subsection{Small scale clustering in real-space}
The continuity equation argument is restricted to the large
scales on which linear theory applies.
The virtue of writing this in terms of halo abundances is
that it shows clearly how to extend the model to predict the
clustering signal in the no-merger model even on small scales.
In particular, two additional pieces of information are required:
a model for how the galaxies are distributed around the centre
of their parent haloes, and the second factorial moment $G_2(M)$ 
of the distribution $p(N|M)$ of the number of galaxies $N$ at fixed 
halo mass $M$.
Sheth et al. (2001) show that, on scales larger than approximately half a
Megaparsec, it is more important to model the first two moments 
$G_1(M)$ and $G_2(M)$ accurately than the density profiles; in particular,
the approximation that the spatial distribution of the galaxies is the 
same as that of the dark matter is sufficiently accurate.  
Hence, if we know the second factorial moment of how galaxies
populate haloes, then we can describe the no-merger correlation
function on small scales as well.
Simulations indicate that 
in haloes which host more than one
galaxy, $p(N-1|M)$ is a Poisson distribution with mean $G_1-1$.  
This specifies $G_2(M)$.

\label{lastpage}


\begin{thebibliography}{99}
\bibitem[\protect\citeauthoryear{Abbas 
\& Sheth}{2005}]{2005MNRAS.364.1327A} Abbas U., Sheth R.~K., 2005, MNRAS, 364, 1327
\bibitem[Adelman-McCarthy et al.(2006)]{2006ApJS..162...38A} 
Adelman-McCarthy, J.~K., et al.\ 2006, ApJS, 162, 38
\bibitem[Adelman-McCarthy et al.(2007)]{2007ApJS..172..634A} 
Adelman-McCarthy, J.~K., et al.\ 2007, ApJS, 172, 634 
\bibitem[Almeida et al.(2008)]{2007arXiv0710.3557A} Almeida, C., Baugh, 
C.~M., Wake, D.~A., Lacey, C.~G., Benson, A.~J., Bower, R.~G., 
\& Pimbblet, K.\ 2008, MNRAS, 386, 2145
\bibitem[Baugh(2006)]{2006RPPh...69.3101B} Baugh, C.~M.\ 2006, Reports of 
Progress in Physics, 69, 3101
\bibitem[Bell et al.(2006)]{2006ApJ...652..270B} Bell, E.~F., Phleps, S., 
Somerville, R.~S., Wolf, C., Borch, A., \& Meisenheimer, K.\ 2006, ApJ, 
652, 270 
\bibitem[Bernardi et al.(1998)]{1998ApJ...508L.143B} Bernardi, M., Renzini, 
A., da Costa, L.~N., Wegner, G., Alonso, M.~V., Pellegrini, P.~S., 
Rit{\'e}, C., \& Willmer, C.~N.~A.\ 1998, ApJL, 508, L143 
\bibitem[\protect\citeauthoryear{Blake, Collister, 
\& Lahav}{2008}]{2008MNRAS.tmp..345B} Blake C., Collister A., Lahav O., 2008, MNRAS, 345, 1257
\bibitem[Bower et al.(2006)]{2006MNRAS.370..645B} Bower, R.~G., Benson, 
A.~J., Malbon, R., Helly, J.~C., Frenk, C.~S., Baugh, C.~M., Cole, S., \& Lacey, C.~G.\ 2006, MNRAS, 370, 645
\bibitem[Brown et al.(2007)]{2007ApJ...654..858B} Brown, M.~J.~I., Dey, A., 
Jannuzi, B.~T., Brand, K., Benson, A.~J., Brodwin, M., Croton, D.~J., \& Eisenhardt, P.~R.\ 2007, ApJ, 654, 858
\bibitem[\protect\citeauthoryear{Bullock et 
al.}{2001}]{2001MNRAS.321..559B} Bullock J.~S., Kolatt T.~S., Sigad Y., 
Somerville R.~S., Kravtsov A.~V., Klypin A.~A., Primack J.~R., Dekel A., 
2001, MNRAS, 321, 559
\bibitem[Cannon et al.(2006)]{2006MNRAS.372..425C} Cannon, R., et al.\ 2006, MNRAS, 372, 425
\bibitem[Casas-Miranda et al.(2002)]{2002MNRAS.333..730C} Casas-Miranda, 
R., Mo, H.~J., Sheth, R.~K., \& Boerner, G.\ 2002, MNRAS, 333, 730
\bibitem[\protect\citeauthoryear{Conroy, Ho, 
\& White}{2007}]{2007MNRAS.379.1491C} Conroy C., Ho S., White M., 2007, MNRAS, 379, 1491
\bibitem[Cooray \& Sheth(2002)]{2002PhR...372....1C} Cooray, A., \& Sheth, R.\ 2002, Phys. Rep., 372, 1
\bibitem[Croton et al.(2006)]{2006MNRAS.365...11C} Croton, D.~J., et al.\ 
2006, MNRAS, 365, 11 
\bibitem[Davis \& Peebles(1983)]{1983ApJ...267..465D} Davis, M., \& 
Peebles, P.~J.~E.\ 1983, ApJ, 267, 465
\bibitem[De Lucia et al.(2006)]{2006MNRAS.366..499D} De Lucia, G., 
Springel, V., White, S.~D.~M., Croton, D., \& Kauffmann, G.\ 2006, MNRAS, 366, 499
\bibitem[Eisenstein et al.(2001)]{2001AJ....122.2267E} Eisenstein, D.~J., 
et al.\ 2001, AJ, 122, 2267 
\bibitem[Font et al.(2008)]{Font2008} Font, A.~S., et al.\ 2008, MNRAS, submitted
\bibitem[Fry(1996)]{1996ApJ...461L..65F} Fry, J.~N.\ 1996, ApJL, 461, L65
\bibitem[Fukugita et al.(1996)]{1996AJ....111.1748F} Fukugita, M., 
Ichikawa, T., Gunn, J.~E., Doi, M., Shimasaku, K., \& Schneider, D.~P.\ 
1996, AJ, 111, 1748  
\bibitem[Gardner et al.(2007)]{2007arXiv0709.1967G} Gardner, J.~P., 
Connolly, A., \& McBride, C.\ 2007, ArXiv e-prints, 709, arXiv:0709.1967
\bibitem[\protect\citeauthoryear{Ho et al.}{2007}]{2007arXiv0706.0727H} Ho 
S., Lin Y.-T., Spergel D., Hirata C.~M., 2007, arXiv, 706, arXiv:0706.0727
\bibitem[Hopkins et al.(2006)]{2006ApJS..163....1H} Hopkins, P.~F., 
Hernquist, L., Cox, T.~J., Di Matteo, T., Robertson, B., \& Springel, V.\ 2006, ApJS, 163, 1
\bibitem[\protect\citeauthoryear{Kaiser}{1987}]{1987MNRAS.227....1K} Kaiser 
N., 1987, MNRAS, 227, 1
\bibitem[Kravtsov et al.(2004)]{2004ApJ...609...35K} Kravtsov, A.~V., 
Berlind, A.~A., Wechsler, R.~H., Klypin, A.~A., Gottl{\"o}ber, S., Allgood, 
B., \& Primack, J.~R.\ 2004, ApJ, 609, 35
\bibitem[Landy \& Szalay(1993)]{1993ApJ...412...64L} Landy, S.~D., \& Szalay, A.~S.\ 1993, ApJ, 412, 64
\bibitem[\protect\citeauthoryear{Lacey 
\& Cole}{1993}]{1993MNRAS.262..627L} Lacey C., Cole S., 1993, MNRAS, 262, 627
\bibitem[Mandelbaum et al.(2005)]{2005MNRAS.361.1287M} Mandelbaum, R., et 
al.\ 2005, MNRAS, 361, 1287 
\bibitem[Masjedi et al.(2006)]{2006ApJ...644...54M} Masjedi, M., et al.\ 
2006, ApJ, 644, 54
\bibitem[Masjedi et al.(2007)]{2007arXiv0708.3240M} Masjedi, M., Hogg, D.~W., \& Blanton, M.~R.\ 2007, ArXiv e-prints, 708, arXiv:0708.3240
\bibitem[McIntosh et al.(2007)]{2007arXiv0710.2157M} McIntosh, D.~H., Guo, 
Y., Hertzberg, J., Katz, N., Mo, H.~J., van den Bosch, F.~C., 
\& Yang, X.\ 2007, ArXiv e-prints, 710, arXiv:0710.2157
\bibitem[Mo \& White(1996)]{1996MNRAS.282..347M} Mo, H.~J., \& White, S.~D.~M.\ 1996, MNRAS, 282, 347
\bibitem[\protect\citeauthoryear{Navarro, Frenk, 
\& White}{1996}]{1996ApJ...462..563N} Navarro J.~F., Frenk C.~S., White S.~D.~M., 1996, ApJ, 462, 563 
\bibitem[Norberg et al.(2002)]{2002MNRAS.332..827N} Norberg, P., et al.\ 2002, MNRAS, 332, 827 
\bibitem[\protect\citeauthoryear{Nusser 
\& Davis}{1994}]{1994ApJ...421L...1N} Nusser A., Davis M., 1994, ApJ, 421, L1
\bibitem[\protect\citeauthoryear{Oke 
\& Gunn}{1983}]{1983ApJ...266..713O} Oke J.~B., Gunn J.~E., 1983, ApJ, 266, 713
\bibitem[\protect\citeauthoryear{Press 
\& Schechter}{1974}]{1974ApJ...187..425P} Press W.~H., Schechter P., 1974, ApJ, 187, 425
\bibitem[Ross et al.(2007)]{2007MNRAS.381..573R} Ross, N.~P., et al.\ 2007, MNRAS, 381, 573
\bibitem[Saunders et al.(1992)]{1992MNRAS.258..134S} Saunders, W., 
Rowan-Robinson, M., \& Lawrence, A.\ 1992, MNRAS, 258, 134 
\bibitem[Schlegel et al.(1998)]{1998ApJ...500..525S} Schlegel, D.~J., 
Finkbeiner, D.~P., \& Davis, M.\ 1998, ApJ, 500, 525 
\bibitem[Scranton et al.(2002)]{2002ApJ...579...48S} Scranton, R., et al.\ 
2002, ApJ, 579, 48 
\bibitem[Scranton et al.(2005)]{2005scranton} Scranton, R., et al.\ 
1005, astro-ph/0508564
\bibitem[Seo et al.(2007)]{2007arXiv0712.1643S} Seo, H.-J., Eisenstein, 
D.~J., \& Zehavi, I.\ 2007, ArXiv e-prints, 712, arXiv:0712.1643
\bibitem[\protect\citeauthoryear{Sheth et al.}{2001}]{2001MNRAS.326..463S} 
Sheth R.~K., Diaferio A., Hui L., Scoccimarro R., 2001, MNRAS, 326, 463
\bibitem[Sheth \& Lemson(1999)]{1999MNRAS.304..767S} Sheth, R.~K., \& Lemson, G.\ 1999, MNRAS, 304, 767
\bibitem[Sheth \& Tormen(1999)]{1999MNRAS.308..119S} Sheth, R.~K., \& Tormen, G.\ 1999, MNRAS, 308, 119
\bibitem[Sheth \& Tormen(2002)]{2002MNRAS.329..61S} Sheth, R.~K., \& Tormen, G.\ 2002, MNRAS, 329, 61
\bibitem[Skibba, Sheth \& Martino(2007)]{2007MNRAS.382.1940S} Skibba, R. A., Sheth, R.~K., \& Martino, M. C.\ 2007, MNRAS, 382, 1940
\bibitem[\protect\citeauthoryear{Stott et al.}{2007}]{2007ApJ...661...95S} 
Stott J.~P., Smail I., Edge A.~C., Ebeling H., Smith G.~P., Kneib J.-P., 
Pimbblet K.~A., 2007, ApJ, 661, 95
\bibitem[Strauss et al.(2002)]{2002AJ....124.1810S} Strauss, M.~A., et al.\ 
2002, AJ, 124, 1810 
\bibitem[Tinker et al.(2007)]{2007ApJ...659..877T} Tinker, J.~L., Norberg, P., Weinberg, D.~H., \& Warren, M.~S.\ 2007, ApJ, 659, 877
\bibitem[Wake et al.(2006)]{2006MNRAS.372..537W} Wake, D.~A., et al.\ 2006, MNRAS, 372, 537 (Paper I)
\bibitem[White et al.(2007)]{2007ApJ...655L..69W} White, M., Zheng, Z., Brown, M.~J.~I., Dey, A., \& Jannuzi, B.~T.\ 2007, ApJL, 655, L69
\bibitem[York et al.(2000)]{2000AJ....120.1579Y} York, D.~G., et al.\ 2000, 
AJ, 120, 1579
\bibitem[Zehavi et al.(2002)]{2002ApJ...571..172Z} Zehavi, I., et al.\ 2002, ApJ, 571, 172
\bibitem[Zehavi et al.(2005)]{2005ApJ...621...22Z} Zehavi, I., et al.\ 2005, ApJ, 621, 22




\end{thebibliography}
\end{document}